 \documentstyle[12pt,equation]{article}
 \advance \voffset by 1.1 cm
 \advance \voffset by -0.8 cm
 \advance \hoffset by 0.5 cm
 
\begin{document}
 \def\ub{\underbar}
 \def\phad{\mbox{$P_{\rm had}$}}
 \def\nav{\mbox{$\langle n_{\rm jet} \rangle $}}
 \def\wgg{\mbox{$W_{\gamma\gamma}$}}
 \def\gga{\mbox{$G^{\gamma}$}}
 \def\qqbar{\mbox{$q \bar{q}$}}
 \def\bbbar{\mbox{$b \bar{b}$}}
 \def\ccbar{\mbox{$c \bar{c}$}}
 \def\ttbar{\mbox{$t \bar{t}$}}
 \def\be{\begin{equation}}
 \def\ee{\end{equation}}
 \def\bea{\begin{eqnarray}}
 \def\eea{\end{eqnarray}}
 \def\qsq{\mbox{$Q^2$}}
 \def\xgam{$ x_\gamma$}
 \def\sigtot{${\sigma^{tot} _W}$}
 \def\gamg{${ \gamma g }$}
 \def\pbarp{$ \bar p p$}
 \def\ptmin{\mbox{$ p_{T,min} $}}
 \def\gamgam{\mbox{$\gamma \gamma $}}
 \def\eplem{\mbox{$e^+ e^- $}}
 \def\alphas{\mbox{$ \alpha_s$ }}
 \def\rts{\mbox{$ \sqrt{s} $ }}
 \def\gamp{${ \gamma p}$}
 \def\totwo{$2 \rightarrow 2$}
 \def\glph{\mbox{${ G^{\gamma}(x,Q^2)}$}}
 \def\qvph{\mbox{${ \vec q^{\gamma} } $}}
 \def\qph{\mbox{${q^{\gamma} } $}}
 \def\vqxqsq{\mbox{$ \vec q^{\gamma} (x,Q^2)  $}}
 \def\vqisq{\mbox{${ q_i^{\gamma} (x,Q^2) }$}}
 \def\vqisqm{\mbox{$ q_i^{\gamma} (x,Q^2)$ }}
 \def\qolsq{ (Q^2 / \Lambda ^2) }
 \def\vqx{{ \vec q^{\gamma} \big( {{x_m} \over x},Q^2 \big) } }
 \def\vqz{{ \vec q^{\gamma} \big ({{z_m} \over z},Q^2 \big ) }}
 \def\fgme{\mbox{$ f_{\gamma/e}$ }}
 \def\fgmebeam{\mbox{$ f_{\gamma/e}^{beam}$ }}
 \def\fgmer{\mbox{$ f_{\gamma/e}^{res} $}}
 \def\lnqmes{ \ln \big({{Q^2} \over {m_e^2}} \big)}
 \def\ebarx{ (\bar E,x) }
 \def\ebarz{ (\bar E,z) }
 \def\ebarzn{${ (\bar E,z)}$}
 \def\fgmen{${ f_{\gamma|e} } $}

 \def\sigh{ \hat \sigma}
 \def\xtsq{\mbox{$ x{_T}{^2}$}}
 \def\alpas{\mbox{${ \alpha_s}$}}
 \def\f2gam{\mbox{$ F_2^\gamma $}}
 \renewcommand{\thefootnote}{\fnsymbol{footnote}}
 \setcounter{page}{0}
 \setcounter{footnote}{0}
 \noindent
 BU-TH-92/5\\
 TIFR/TH/92--70\\
 Dec. 1992\\
 \vspace{1cm}
 \begin{center}
   \begin{Large}
   \begin{bf}
        Resolved Photon Processes
 \\
   \end{bf}
   \end{Large}
   \vspace{1cm}
               {\large  Manuel Drees$^1$\footnote{Address after February 1993:
 University of Wisconsin, Dept. of Physics, 1150 University Ave., Madison,
 WI53706, USA} and Rohini M. Godbole$^2$\\}
        $^1$ Tata Institute of Fundamental Research, Homi Bhabha Road, Bombay
      400 005, India.\\
        $^2$ Physics Department, University of Bombay, Vidyanagari, Bombay,
          400 098, India.\\
  \vspace{1cm}
 \end{center}
 \begin{abstract}
 We review high--energy scattering processes that are sensitive to the hadronic
 structure of the photon, describing theoretical predictions as well as
 recent experimental results. These processes include deep--inelastic
 electron--photon scattering at \eplem\ colliders; and the production of jets,
 heavy quarks and isolated photons in the collision of real photons at \eplem\
 colliders, as well as in photon--proton collisions at $ep$ colliders. We also
 comment on ``minijet'' based calculations of total $\gamma p$ and
 \gamgam\ cross--sections, and discuss the possibility that future linear
 \eplem\ colliders might produce very large photon fluxes due to the
 ``beamstrahlung'' phenomenon; in the most extreme cases, we predict more
 than one hadronic \gamgam\ event to occur at every bunch crossing.
 \end{abstract}
 \newpage
 \pagestyle{plain}
 \section*{1) Introduction}
 \setcounter{footnote}{0}
 Among the quarks, leptons and gauge bosons now thought to be truly
 elementary particles, the photon occupies a special place. Together with the
 electron it was the first elementary particle correctly identified as such;
the understanding of reactions involving these two particles spawned the theory
 of gauge interactions, now thought to describe all observed (electroweak,
 strong and gravitational) interactions. In view of this long, distinguished
 history, it may come as a surprise that there is one large class of photonic
 interactions about which only relatively little is known: The interaction
 of real (on--shell) photons with hadrons (or other real photons).

 Such reactions can proceed in two quite different ways: The photon can
 couple directly to a quark or gluon in the struck hadron; in this case
 the whole energy of the photon goes into the hard (partonic) scattering
 process. Alternatively, the photon can undergo a transition into a
 (virtual) hadronic state before encountering the target hadron. In this
 case a quark or gluon ``in'' the photon can react, via strong interactions,
 with partons in the struck hadron. Notice that now only a fraction of the
 photon's energy goes into the partonic scattering; the rest is carried
 away by a ``spectator jet'' produced by the break--up of the photon. (In
 both cases the break--up of the struck hadron will also produce such a
 spectator or remnant jet, at least in the photon--hadron centre--of--mass
 frame.) In this article we generally refer to the first kind of reaction
 as {\em direct processes}, while those of the second kind are termed
 {\em resolved photon processes}. Cross sections for direct processes are
 computable from perturbative QCD (assuming the reaction is ``hard''
 enough, i.e. involves a sufficiently large momentum exchange) in terms of
 parton densities inside the hadron. Similarly, the cross sections for
 resolved photon processes depend on the parton densities ``inside'' the
 photon.

 Since the pioneering SLAC measurement \cite{slac} of deep--inelastic
 electron--nucleon scattering, a large body of data on the parton
 distributions inside nucleons has been accumulated \cite{nucrev}.
 Further constraints on nucleonic parton densities are imposed by several sum
 rules, which can be derived directly from QCD. In comparison the picture looks
 much more sketchy where the parton content of the photon is concerned. Until
 very recently, the only relevant data were measurements of the electromagnetic
 structure function \f2gam\ in deep--inelastic $e \gamma$ scattering; as we
 will see in more detail in sec. 2, these measurements suffer from large
 theoretical and/or experimental uncertainties. Moreover, they only cover a
 limited kinematical range. Finally, they are not sensitive to the gluon
 content of the photon, which plays an important role in many resolved
 photon processes. During the last year experimental analyses of jet
 production in \gamgam\  collisions in
 terms of resolved photon processes have started to appear \cite{amy,sandiego}.
 Even more  recently, the $ep$ collider HERA has started operations;
 photoproduction processes, including resolved photon reactions, play an
 important role in its physics programme \cite{zeus,h1}. It seems therefore
 timely to review  what is known
 about resolved photon processes and the parton content of the photon, and
 what we can hope to learn in the near future. We will see in sec. 5 that
 this information might be crucial for assessing the physics potential of
 future linear \eplem\ supercolliders.

 This article is organized as follows. In sec. 2 we give a short introduction
 to the photon structure function, describing the theoretical understanding of
 as well as experimental data on \f2gam. We also briefly compare existing
 parametrizations of photonic parton densities. In sec. 3 we discuss hard
 \gamgam\ processes at existing and planned \eplem\ colliders, while sec. 4
 is devoted to hard $\gamma p$ reactions. We will see in both cases that the
 relative importance of resolved photon processes steadily increases with the
 available centre--of--mass energy. As a result, at existing \eplem\
 colliders one can probably only study processes with large partonic cross
 sections, i.e. jet and heavy quark production; in contrast, at HERA and
 certain \eplem\ supercolliders a large variety of final states that receive
 contributions from resolved photon processes can be produced with detectable
 rates. In sec. 5 we discuss to what extent ``minijet'' calculations allow
 us to predict total $\gamma p$ and \gamgam\ cross sections. We will show that
 even if the total $pp$ cross section and the parton densities in the
 photon were known, sizeble theoretical uncertainties would remain due to
 our insufficient understanding of multiple parton scattering. Finally, sec. 6
 contains a summary of our results.

 \renewcommand{\theequation}{2.\arabic{equation}}
 \section*{2) The Photon Structure Function $F_2^\gamma $}
 \setcounter{footnote}{0}
 As we know, information about the proton structure is obtained by
 studying Deep Inelastic Scattering (DIS) of high energy leptons of
 energy E off proton targets,
 \be
 e^- + p \rightarrow e^- + X
 \label{dis}
 \ee
 shown in
 \begin{figure}[hbt]
 \vspace{4cm}
 \caption{Deep Inelastic Scattering for the proton and photon.}
 \label{disfig}
 \end{figure}
 fig. \ref{disfig}a. The structure of the proton as revealed to a
 photon probe of invariant mass $ - Q^2 $ depends on the value of
 \qsq. In the DIS regime the process of eq.(\ref{dis}) is
 characterised by two independent kinematic variables, $y = \nu /E $  where
 $\nu $ is the energy carried by the probing photon in the laboratory
 frame, and $x=Q^2 /( 2 M \nu ) $ where M is the proton
 mass. We also know that the double differential cross--section for
 this process  factorises in the quark-parton model (QPM) as,
 \be
 \frac{d^2\sigma^{ep \rightarrow X}}  {dx dy} = \frac {2 \pi\
 \alpha^2 \ s}
 {Q^4} \times \bigg[(1+(1-y)^2)\ F_2^p (x) - y^2\ F_L^p(x) \bigg],
 \label{discs}
 \ee
 where
 \bea
  F_2^p(x)& = &  \sum_{q} e_q^2 \  x\ q^P(x); \nonumber \\
  F_L^p(x) & = &  F_2^p(x) - 2xF_1^p (x)
 \eea
are the two electromagnetic structure functions of the proton. The longitudinal
 structure function $F_L^p(x) $ is zero in QPM, $ q^p(x) $  the
 probability for quark $q$ to carry a momentum fraction $x$ of the proton and
 $e_q$ denotes the electromagnetic charge of quark $q$ in units of the
 proton charge. This factorisation of the $x$ and \qsq\ dependence in
 eq.(\ref{discs}) is of course only approximate. In general $F_2^p$
 depends on $x$ as well as \qsq\ and $ F_L^p(x) $ is nonzero. QCD
 predicts the \qsq\ dependence  of the structure functions, given by the
 Gribov--Lipatov--Altarelli--Parisi (GLAP) \cite{glap}
 equations but does not say anything about the shape
 of hadronic structure functions.
 According to QCD predictions all the hadronic structure functions
 shrink to lower values of $x$ as \qsq\ increases.

 The idea that photons behave like hadrons when interacting with
 other hadrons dates back to the early days of strong interaction physics
 and is known to us under the name of the Vector Meson Dominance (VMD)
 picture. This essentially means that at low 4--momentum transfer,
 the interaction of a
 photon with hadrons is dominated by the exchange of vector mesons which
 have the same quantum numbers as the photon.  While this
 picture works reasonably well for ``soft'' processes (i.e., reactions
 characterized by small 4--momentum transfer), it is not at all
 clear that it should describe the whole story of interactions of
 photons with hadrons at high energies as well.

 Since a photon is known to behave like a hadron, it seems reasonable
 that it should be possible to probe its structure also in a DIS
 experiment. Such an experimental situation is provided at \eplem\
 colliders in $\gamma^* \gamma $ reactions as shown in
 fig.~\ref{disfig} (b). Here the virtual photon with invariant mass square
 $-Q^2 $ probes the structure of the real photon. If the VMD
 picture were the whole story then one would expect that such an
 experiment will find
 \be
 F_2^\gamma \simeq F_2^{\gamma,VMD} \propto F_2^{\rho^0} \simeq F_2^{\pi^0}.
 \label{f2VMD}
 \ee
 Then with increasing \qsq,
 the  structure function \f2gam\ will behave just like a hadronic
 proton structure function. However, there is a very important
 difference in case of photons,{\it i.e.}, photons possess pointlike
 couplings to quarks. This has interesting implications for $\gamma^* \gamma $
 interactions as first noted in the framework of the
 QPM by Walsh \cite{wal}.  It
 essentially means that $\gamma^* \gamma $ scattering in
 fig.~\ref{disfig} contains two contributions as shown in
 \begin{figure}[hbt]
 \vspace{4.5cm}
 \caption{Two contributions to \f2gam.}
 \label{fgamtwo}
 \end{figure}
 fig.~\ref{fgamtwo}. The contribution of fig.~\ref{fgamtwo}~(a) can be
 estimated by eq.(\ref{f2VMD}), whereas that of fig.~\ref{fgamtwo}~(b)
 was calculated in the QPM~\cite{wal}. This is done by considering
 the cross--section for the reaction
 \begin{displaymath}
 \gamma + \gamma^* \rightarrow q + \bar q.
 \end{displaymath}
Due to $t$ and $u$ channel poles this can be calculated only when one considers
 quarks with finite masses. The result can  be recast in a form equivalent to
 eq.(\ref{discs}):
 \bea
 { {d^2\sigma^{e\gamma \rightarrow X}} \over {dx dy}} &= &{ {2
 \pi
 \alpha^2  s_{e \gamma}} \over {Q^4}} \times {3 \alpha \over \pi}
 \nonumber \\
 & & {  \sum_q e_q^4 \bigg\{ (1+(1-y)^2)\times {\bf [}x(x^2
 +(1-x)^2) \times \ln{{W^2} \over {m_q^2}}} \nonumber \\
 & & +{ 8x^2 (1-x) -x{\bf ]} - y^2 {\bf [} 4 x^2 (1-x)
 {\bf ]} \bigg\}.},
 \label{qpm}
 \eea
 where $ W^2 = Q^2 (1-x)/ x.$
 On comparing Eqs.(\ref{discs}) and (\ref{qpm}) we see that the
 factors in square brackets in the above equation  have the
 natural interpretation as photon structure functions $ F_2^\gamma $
 and $F_L^\gamma $ and one has
 \bea
 F_2^{\gamma,\rm{pointlike}}(x,Q^2) &= & 3 {\alpha \over \pi}
                             { \sum_q e_q^4 \bigg[ x (x^2 +(1-x)^2)
                              \times
 \ln{{W^2} \over {m_q^2}} + 8x^2(1-x) -x \bigg]} \nonumber \\
                        & = & {\sum_q e_q^2  x\
                        q^{\gamma,\rm{pointlike}}(x,Q^2)}.
 \label{qpmpred}
 \eea
 Two points are worth noting:  the function $
 F_2^{\gamma,\rm{pointlike}}(x,Q^2) $ can be completely calculated in QED
 and secondly this contribution to \f2gam\ increases logarithmically
 with \qsq. So in this simple ``VMD + QPM'' picture, \f2gam\ consists of
 two parts, $F_2^{\gamma,\rm{pointlike}}$ and $ F_2^{\gamma,\rm{VMD}}$, with
 distinctly  different \qsq\ behaviour and with the distinction that
 for one part both the $x$  and the \qsq\  dependence can be calculated
 completely from first principles.

 This QPM prediction received further support when it was shown by
 Witten \cite{wit77} that at large \qsq\ and at large $x$, both the $x$
 and \qsq\ dependence of the quark and gluon densities in the photon can
 be predicted completely even after QCD radiation is included.  An alternative
 way of understanding this result is to consider the evolution
 equations \cite{dewit} for the  quark and gluon densities
 inside the photon. These contain an inhomogeneous term on  the r.h.s
 proportional to $\alpha$, which describes $\gamma \rightarrow q \bar{q}$
 splitting, i.e. the pointlike coupling of photons to quarks. In the
 `asymptotic' limit of large \qsq\ and large $x$, the  \vqisq have
 the form
 \bea
 q_i^{\gamma ,\rm{asymp}}(x,Q^2)& \propto & {\alpha \times
 \ln \left( {Q^2}
 \over {\Lambda^2_{\rm{QCD}}} \right) F_i(x) }\nonumber \\
 &\simeq & {{\alpha \over {\alpha_s}} F_i(x)},
 \label{asymp}
 \eea
 where $\rm \Lambda_{QCD} $ is the usual QCD scale parameter and the $x$
 dependence of the $F_i(x)$ is completely calculable. If we compare
 eqs. (\ref{qpmpred}) and (\ref{asymp}) we see that the essential
 change, apart from the different $x$ dependence, between
 the leading order (LO) QCD and QPM predictions is the replacement of $m_q^2
 \rightarrow \Lambda^2.$ We then have
 \be
 F_2^{\gamma,\rm{asymp}}  = \sum_q e^2_q x q^{\gamma,\rm{asymp}}
 \ee
 The asymptotic form of \glph\ can also be uniquely calculated.
 These asymptotic predictions, however, show unphysical
 divergences as $x \rightarrow 0$ indicating thereby that at
 small $x$ the ``hadronic'' part of \f2gam\ can not be neglected, and is not
 adequately described by a regular, VMD--inspired ansatz.
 The initial enthusiasm that this
 completely calculable prediction can be used as a test of QCD and
 also for a high precision measurement of \alpas, suffered a big set
 back by the observation \cite{2loop} that the degree of the $x \rightarrow 0$
 pole becomes larger in higher orders of perturbation theory; we therefore
 have to conclude that the
 separation of \f2gam\ in two parts $ F_2^{\gamma,\rm{VMD}} $ and
 $F_2^{\gamma,\rm{asymp}} $ is not physical.  One way out of this is to add
 a hadronic contribution to $F_2^{\gamma,\rm{asymp}}$ with similar
 divergence so as to get a finite result at small $x$ \cite{anto}. This
 involves  arbitrary parameters, but keeps the hope of being
 able to use the calculable pointlike contribution for QCD tests. An
 alternative suggestion is to use the experimental data on \f2gam\ to fit
 a finite  input distribution for \qvph\ at a scale  $Q_0^2 \simeq 1\
 \rm{GeV}^2 $, thus retaining the  predictability only of the  \qsq\
 dependence but gaining the ability to calculate physically meaningful
 quantities for all combinations of $x$ and $Q^2 \geq Q_0^2$ \cite{gr}.
 The theoretical debate on the subject is still
 not completely closed \cite{debate}.

 By now several parametrizations of the parton densities inside the photon
 $\vqxqsq \equiv (q_i^{\gamma},\gga)(x,Q^2)$ exist. The oldest parametrizations
 \cite{do,nico} are based on the ``asymptotic'' LO predictions. However,
 these parametrizations are even more singular as $x \rightarrow 0$ than the
 exact ``asymptotic'' prediction. Care must therefore be exercised if these
 parametrizations are to be used at small $x$.

 The first parametrization that followed the suggestion of ref.\cite{gr} to
 fit input distributions at some low scale $Q_0^2$ such that data at higher
 $Q^2$ are reproduced is the ``DG'' parametrization of ref.\cite{dg}. In 1984,
 when this parametrization was constructed, only a single set of data on
 \f2gam\ existed \cite{plut84}. Since then, more data have become available
 \cite{exrev,sandiego}. Most of these newer data were taken into account in
 the ``LAC'' fits of ref.\cite{lac}. Indeed, taken at face value, low$-Q^2$
 data from the TPC/2$\gamma$ collaboration \cite{tpc} at PEP disfavour the
 older DG parametrization compared to the LAC fits. However, the interpretation
 of these data is not entirely straightforward. A general problem of the
measurement of the $x$ dependence of \f2gam\ is that, unlike in deep--inelastic
 $ep$ scattering, $x$ has to be determined from the hadronic final state, since
 the energy of the target photon is not known\footnote{Recall that in the
 ``single--tag'' experiments we are discussing here, one of the electrons
 emerges at too small an angle to be detected.}. Unfortunately some of the
 final state particles are usually lost in the beam pipe. One therefore needs
 fairly sophisticated ``unfolding'' techniques in order to determine the true
 value of $x$ from the observed final state. Notice that some model of
 \f2gam\ has to serve as an {\em input} for this unfolding procedure. Usually
 a simple ``QPM+VMD'' description is used for this purpose, even though we
 have seen above that this picture is not very meaningful within QCD.

 Low$-Q^2$ data suffer from two additional problems. First of all, higher twist
 contributions might be quite important. Often only the region $W \geq 2$ GeV
 is used for QCD comparisons, i.e. the region of large $x$ is discarded; it is
 not clear, however, whether this is sufficient to really make higher twist
 contributions negligibly small. Secondly, at low $x$ and low $Q^2$ one
 actually has $W^2 \gg Q^2$; in this case it is not clear that $Q^2$ is
 indeed the relevant scale in the process. Ideally one would like a
 unified treatment of real \gamgam\ (no--tag) and $\gamma^*\gamma$
(single--tag)
 data, allowing for a smooth transition from one to the other. Work along
 these lines is in progress \cite{frederic}. At present we do not think it
 advisable to base the exclusion of a parametrization of \vqxqsq\  on these
 low$-Q^2$ data alone.

 The paramerizations of refs.\cite{dg,lac} are the only ones that allow an
 arbitrary form at least of the quark densities at scale $Q_0^2$; this leads
 to a fairly large number of free parameters which, in view of the paucity
 of good data, cannot be determined very precisely. The authors of
 ref.\cite{grv} have therefore used theoretical considerations (or prejudices)
 to fix the shape of the input distributions, reducing the number of free fit
 parameters to one. They assume that eq.(\ref{f2VMD}) is valid at the rather
 low input scale $Q_0^2 = 0.25\ \rm{GeV}^2$; this scale has been taken from
 the ``dynamical'' fits to proton structure functions by the same authors
 \cite{grvn}. Similarly, their pionic structure functions \cite{grvpi}
 were used to fix the shape of the input distributions. The only free
 parameter is then the overall size of the input distributions; the idea is
 that this parameter describes the contribution of vector mesons beyond
 the $\rho$.

 Finally, for the ``GS'' parametrization of ref.\cite{gs} an intermediate
 approach has been taken. In
 order to avoid the ambiguities of low$-Q^2$ data, a rather high input
 scale $Q_0^2$ = 5.3 GeV$^2 $ has been chosen. At this scale, \f2gam\
 is assumed to be described by the ``QPM+VMD'' model; however, the
 gluon and sea--quark distributions in the pion, the overall size of
 the VMD (pionic) contribution, as well as the quark masses in
 eq.(\ref{qpmpred}) are all allowed to vary within reasonable limits.

 In fig. \ref{parcomp}a we compare various parametrizations for
 $\f2gam(x, Q^2 = 5.3$ GeV$^2) $ with each other and with experimental data
from
 the PLUTO collaboration \cite{plut84}. Since only the contributions from
 $u,d$ and $s$ quarks have been included in the calculation, charm--subtracted
 data have been used. We observe that all shown parametrizations describe the
 data reasonably well, although none of the fits is perfect. The differences
 between the various parametrizations only amount to typically 20\% in the
 region $x \geq 0.05$. Due to the inhomogeneous term in the evolution
 equation, these differences tend to be even smaller at higher $Q^2$;
 statistical errors are also larger for these theoretically cleaner high$-Q^2$
 data. The best possibility to discriminate between different parametrizations
 using data on \f2gam\ therefore seems to lie in high--statistics measurements
 at intermediate values of $Q^2$ and small $x$.

 \begin{figure} [hbt]
 \vspace{8cm}
 \caption{Comparison of existing parametrizations of \protect\qvph.
 The data are from \protect\cite{plut84}.}
 \label{parcomp}
 \end{figure}

 In fig. \ref{parcomp}b we compare the parametrizations of the gluon density
 at the same value of $Q^2$. Obviously the various parametrizations differ
 much more strongly here than for the quark densities that determine \f2gam.
 The extreme behaviour of the LAC parametrizations is especially noticeable;
 this is because in these parametrizations no theoretical assumptions about
 $\gga(x, Q_0^2)$ were made. The very hard gluon density of LAC3 is
 necessary to explain the rapid increase of $\f2gam(x\simeq0.1,Q^2\leq4 \
 \rm{GeV}^2)$ with $Q^2$ seen in the data \cite{tpc}; at $Q^2>Q_0^2$, some
 of the gluons at large $x$ are converted into \qqbar\ pairs, leading to
 the maximum of $\f2gam(x\simeq0.1)$ shown in fig. \ref{parcomp}a for this
 parametrization. These low$-Q^2$ data were ignored for the LAC2 fit, which
 uses $Q_0^2 = 4$ GeV$^2$; now a very soft gluon density is favoured. It
 should be noted, however, that the LAC3 parametrization also describes the
 high$-Q^2$ data adequately; this shows that \f2gam\ is not very sensitive
 to \gga.

 For the other three parametrizations the gluon input is essentially fixed
 from the quark input. We already saw that the GRV fit only contains a
 single free parameter, which is fixed from data on \f2gam. In the DG
 parametrization it is assumed that gluons are only produced radiatively
 from quarks, i.e. there is no truly intrinsic (VMD--like or otherwise)
 gluon content of the photon; as a result, the DG gluon density falls below
 that of the other parametrizations, except for very small $x$ where it
 resembles the GRV and LAC3 gluons. Note, however, that the DG parametrization
 has been obtained using a rather large value for the QCD scale parameter
 $\Lambda =
 400$ MeV, while all other parametrizations assume $\Lambda = 200$ MeV.
 The same value of $\Lambda$ should also be used in the factors of $\alpha_s$
 that occur in resolved photon cross--sections. As a result, the DG
parametrization sometimes leads to larger predictions for such cross--sections,
 in spite of its smaller gluon content. Finally, in the GS2 parametrization,
 $\gga(x,Q_0^2)$ receives contributions both from a VMD--like (pionic) term,
 and from radiation off the (QPM) quarks.\footnote{We do not show results for
 the LAC1 and GS1 parametrizations, since they are qualitatively similar to
 the LAC2 and GRV parametrizations, respectively.}

 There are also versions of the GRV and GS parametrizations that include
 higher order effects, i.e. where the 2--loop evolution equations
 \cite{2loop} have been used. However, in view of the large uncertainties
 at least in the gluon densities, this seems at present an uneccessary
 refinement. Moreover, full higher order calculations do not yet exist for
 most resolved photon processes; we will come back to this point later.

 This concludes our discussion of deep--inelastic $e\gamma$ scattering.
 While theoretically relatively clean, since one computes a fully inclusive
 cross section, we have seen that this is probably not the best way to give
 us detailed information about the partonic structure of the photon, as
 witnessed by the large differences between parametrizations of the gluon
 density. As we will see in the following sections, this information might
 be provided by resolved photon processes involving only real photons.

 \renewcommand{\theequation}{3.\arabic{equation}}
 \setcounter{equation}{0}
 \setcounter{footnote}{0}
 \section*{3) Real \gamgam\ Scattering at \eplem\ Colliders}

 Effects of the photon structure and especially the gluon
 component of the photon are best studied in processes involving real
 photons. In this section we will discuss only the possibilities of
 probing the photon structure at \eplem\ colliders. Possible effects
 of the hadronic component of the photon at $ep$ colliders and at
 fixed target photoproduction experiments will be discussed in the
 next section. At \eplem\ colliders processes other than DIS which
 could yield information about the photon structure function are jet
 production, heavy flavour production and prompt photon production
 initiated by partons in the photon. Compared to DIS (single--tag)
 processes, real \gamgam\ reactions have cross sections which are
 enhanced by a factor $(\ln s / (4 m_e^2)) \simeq 20$ at existing
 colliders. The
 scale at which the parton densities in the photon, \vqxqsq, will be
 probed is decided by the $p_T^2 $ of the jets, which is comparable to
 the higher end of $Q^2$ values accessible to DIS experiments at
 present colliders.

 Jet production in \gamgam\ collisions can receive contributions from
 three different types of diagram \cite{llewellyn} as shown in
 \begin{figure}[hbt]
 \vspace{9.5 cm}
 \caption{Different contributions to the production of high $p_T$
 jets in \eplem\  collisions with the associated topologies.}
 \label{ptjts}
 \end{figure}
 fig.~\ref{ptjts}. The `direct process' of fig.~\ref{ptjts}a is due
 to $\gamgam \rightarrow \qqbar\ $  production, present already in the
 naive quark-parton model. Fig.~\ref{ptjts}b depicts the case where
 only one photon is resolved into its partonic components, which then
 interact with the other photon; we call these the `once-resolved'
 processes (`1-res' for short).  Finally, fig.~\ref{ptjts}c shows
 the situation where both photons are resolved, so that the hard
 scattering is a pure QCD \totwo\ process; we call these the
 `twice-resolved' contributions (`2-res' for short). It is very
 important to note here that every resolved photon will produce a
 spectator jet of hadrons with small transverse momentum relative to
 the initial photon direction, which for (quasi-~) real photons
 coincides with the beam direction. The resolved contributions of
 fig.~\ref{ptjts}b and c can therefore be separated if one can
 tag on these spectator jets.

 The cross-section for jet production in \gamgam\ collisions for the
 `2-res' processes can be written schematically as \cite{oldjet,tristanac}
 \be
  d\sigma = f_{\gamma/e}(x_1) \vec q^{\gamma} (x_2,Q^2)
           f_{\gamma/e}(x_3) \vec q^{\gamma} (x_4,Q^2) d\sigh,
 \label{csec}
 \ee
 where the $\hat{\sigma}$ are the cross sections for  the hard \totwo\
 subprocesses \cite{comkrip,do}, \vqxqsq, \fgmen\  denote
 parton densities inside the photon and  photon fluxes inside the
 electron respectively; we include non--leading contributions to \fgmen,
 following ref. \cite{bkt}. For the `1-res' (direct)
 processes, one (both) of the parton density functions \vqxqsq\ have
 to be replaced by $ \delta (1-x)$, and the proper hard sub-process
 cross--sections have to be inserted \cite{do}.  Recall
 eq.(\ref{asymp}) for \qvph. This makes it clear that all three classes of
 diagrams are of the same order in $\alpha$ and \alpas.

 While it is clear that our present knowledge of \qvph\ is not precise
 enough to make absolute predictions, one can check how
 sensitively the predicted cross--sections depend on the choice of
 \qvph. The DG parametrisation will usually give us the most conservative
 prediction of the available parametrisations, as can be seen from
 fig. \ref{parcomp}.

 In fig. \ref{csrts} we show the energy dependence of the cross--section
 for the production of two jets with $p_T = 3$ GeV,
 as predicted \cite{tristanac} by the DG parametrization, in the range
 covered by the PETRA and TRISTAN colliders. For this choice of $p_T$ the
 cross--section is quite sizable; recall that the total luminosity
 collected at PETRA amounts to several hundred pb$^{-1}$ per experiment,
 while as of this writing, TRISTAN has collected about 100 pb$^{-1}$. The
 cross--section is also well above the background from annihilation events
 with hard initial state radiation (dotted curve). We also note that the
 twice--resolved contribution grows faster than \rts\ with increasing
 machine energy and, for this choice of $p_T$, begins to dominate the
 cross--sections in the energy range of TRISTAN. Notice that ``jets'' with
 smaller transverse momentum can even originate from the poorly understood
 soft (VMD) contribution to the \gamgam\ cross section; this contribution
 is essentially negligible if $p_T \geq 2$ to 3 GeV. Thus TRISTAN is in a
 unique position to probe the structure of the photon through jet production.
 Our detailed studies \cite{tristanac} do indeed indicate that studying the
 production of high $p_T$ jets and heavy flavour (charm) at TRISTAN  should
 be able to probe the hadronic content of photon in some detail.

 \begin{figure}[hbt]
 \vspace{8 cm}
 \caption{$ d \sigma / d p_T$ at $p_T = 3$ GeV as a
 function of $\protect\sqrt{s}$ \protect\cite{tristanac}.}
 \label{csrts}
 \end{figure}

 This is demonstrated in fig. \ref{tript}, where we compare the $p_T$
 spectrum of jets produced in \gamgam\ collisions at $\rts(\eplem) = 60$
 GeV, as predicted by the DG (a) and LAC3 (b) parametrizations. We already
 saw in the previous figure that the DG parametrization predicts the three
 classes of processes to contribute roughly equally at $p_T = 3$ GeV. At
 larger values of $p_T$ the direct process starts to dominate. The reason is
 simply that a resolved photon has to split its energy between the parton
 participating in the hard scattering on the one hand, and a spectator jet
 on the other; therefore the cross--section for resolved photon processes
 will depend more strongly on the available phase space, i.e. will have a
 steeper $p_T$ spectrum. Nevertheless, even for for the DG parametrization
 the sum of once and twice resolved contributions exceeds the direct
 contribution out to $p_T \simeq 4.5$ GeV.

 \begin{figure}[hbt]
 \vspace{8 cm}
 \caption{The transverse momentum spectrum of jets produced in real
 \protect\gamgam\ scattering at \protect\rts\ = 60 GeV, for the DG (a) and
 LAC3 (b) parametrizations.}
 \label{tript}
 \end{figure}

 The extremely hard gluon density of the LAC3 parametrization (see fig.
 \ref{parcomp}b) greatly enhances the cross--section for twice resolved
 processes compared to the predictions of the DG parametrization.
 Note that out of the eight \totwo\ QCD scattering matrix elements, that for
 $gg \rightarrow gg$ scattering is the largest, followed by the one for
 $qg \rightarrow qg$ \cite{comkrip}. As a result, LAC3 predicts the
 2--res contribution to be almost an order of magnitude larger than DG;
 it also predicts an approximately two times larger 1--res contribution.
 The LAC3 parametrization therefore predicts the high$-p_T$ jet
 cross--section to be dominated by resolved photon contributions up to
 $p_T \simeq 9.5$ GeV.

 What is the experimental situation? Note that even at the lower end of the
 curves in fig. \ref{csrts}, at PEP/PETRA energies, one expects a sizable
 contribution to the jet cross--section from resolved
 processes in addition to the `direct' process.  In this context it is
 interesting to note that almost all the groups at PEP/PETRA
 observed \cite{thirdcomp} such an excess of jet events. These
 experiments compared data with a two component model where they added
 to the direct process a soft component (with exponential $p_T$
 spectrum) expected from the VMD picture
 and always {\em failed} to reproduce the data. The data always had
 yet another `third component' with a $p_T$ spectrum softer than QPM
 but broader than VMD and  thrust distribution broader than the QPM
 prediction. Both features are expected of resolved contributions.

 \begin{figure}
 \vspace {7cm}
 \caption{Data on \gamgam $\rightarrow $ jets and predictions
  with  and without `res' contributions \protect\cite{amydiego}.}
 \label{amydat}
 \end{figure}

 The first experimental analysis including resolved photon contributions has
 recently been performed by the AMY collaboration at TRISTAN \cite{amy}.
 They modelled the three classes of hard contributions to high$-p_T$ jet
 production using eq.(\ref{csec}), but included the full machinery of
 initial and final state parton showers predicted by QCD, as well as
 parton $\rightarrow$ hadron fragmentation. They conclude that inclusion of
 resolved photon processes as predicted using the DG parametrization greatly
 improves the agreement between Monte Carlo predictions and data.

 An example is shown in fig. \ref{amydat} \cite{amydiego}, which shows the
 $p_T$ spectrum of their data sample. It should be noted that, like
 previous analyses of \gamgam\ scattering \cite{thirdcomp}, AMY does not
 use a jet finding algorithm. Rather, the entire event is divided into
 two hemispheres, perpendicular to the thrust axis; the $p_T$ shown in the
 figure is then simply the sum over the transverse momenta of all particles in
 one hemisphere. Notice also that the AMY trigger requires the event to
 contain at least one charged particle with $p_T \geq 1.0$ GeV. This
 suppresses events with very small $p_T$ per hemisphere, and further
 complicates the relation between the partonic and hemispheric transverse
 momentum.

 The agreement between QCD MC predictions (solid histogram) and data (points)
 shown in fig. \ref{amydat} is indeed quite impressive, in particular when
 compared to the prediction of the traditional ``QPM+VMD'' model (dashed
 histogram). It should be noted that the AMY Monte Carlo contains a number
 of free parameters beyond those determining the \vqxqsq. The most
 important one is the cut--off \ptmin, which is the smallest partonic
 transverse momentum allowed in the hard scattering diagrams of fig.
 \ref{ptjts}. Of course, these diagrams diverge badly as $p_T \rightarrow
 0$; QCD does not tell us, however, just how large the partonic $p_T$ has
 to be for its predictions to become trustworthy. AMY thus simply fits
 \ptmin\ from their data with $p_T^{\rm thrust} \geq 1.5$ GeV; they find
 \ptmin\ = 1.6 GeV for the DG parametrization, if only the three light
 flavours of quarks are assumed to be present in the photon. The same
 value of \ptmin\ also leads to a good description of the $p_T^{\rm thrust}$
 spectrum in the theoretically cleaner region beyond 3 GeV. Other free
 parameters include the amount of intrinsic $p_T$ allowed for the
 partons inside the photon, and parameters describing the hadronization
 process. The experimental observables that have been studied so far,
 in particular the thrust distribution, do not seem to depend much on
 the former, but are quite sensitive to the latter. More detailed
 analyses of higher statistics data taken with an upgraded AMY detector
 are now being carried out. However, the existing data are already good
 enough to rule out the LAC3 parametrization; the data are
 clearly incompatible \cite{amydiego} with the huge rate of resolved
 photon events predicted by this parametrization, see fig. \ref{tript}.
 In contrast, the LAC1 or LAC2 parametrizations, with \ptmin\ = 2.0 GeV,
 seem to describe the recent AMY data slightly better than the DG
 parametrization does \cite{amykek}. On the other hand, a toy--model with
 zero gluon content of the photon falls short of the data, even if
 \ptmin\ is allowed to be as small as 1.0 GeV \cite{amy}. Finally,
 by comparing their own data with data taken at the PETRA collider,
 AMY could show that the importance of the resolved photon processes
 increases with energy \cite{amy}, as expected from fig. \ref{csrts}.

 More recently, other experimental groups have also entered the fray.
 In particular, the TOPAZ collaboration at TRISTAN has presented a
 preliminary analysis \cite{topaz} of their data. Compared to the AMY
 detector, TOPAZ has the advantage of a lower trigger threshold, which
 merely requires the presence of 2 charged tracks with $p_T \geq 0.3$ GeV
 in the event; for a given luminosity, this leads to an approximately two
 times larger \gamgam\ data sample than at AMY. This should allow for a more
 detailed study of the region with low and intermediate $p_T$, which is
 however difficult to interpret theoretically. The $p_T$ distribution
 presented by TOPAZ is in good agreement with the AMY result, while the
 thrust distributions seem to differ somewhat.

 Very recently TOPAZ has also presented \cite{topazkek} preliminary results
 of an analysis which, for the first time in \gamgam\ physics, actually
 requires jets to be reconstructed, using an algorithm familiar from
 hadron collider studies of jets. They used an unfolding procedure to
 extract the {\em partonic} cross--section from the measured jet rates. The
 results
 for the cross--section integrated over 2.5 GeV $\leq p_T$(parton)$\leq$ 8
 GeV are:
 \be
 \begin{eqalign} \label{topaz}
 \sigma (|y_1|, |y_2| \leq 0.7) &= 23.4 \pm 2.7 \pm 1.7 \ {\rm pb} \nonumber \\
 \sigma (|y_1| \leq 0.7, y_2 \ {\rm anywhere}) &= 96.7 \pm 3.7 \pm 8.5 \
 {\rm pb}
 \end{eqalign}
 \ee
 The first result corresponds to the situation where both high$-p_T$ jets
 are reconstructed, while the second result includes events where only one
 high$-p_T$ jet is seen.
 Notice that these are \eplem, not \gamgam, cross--sections. The systematic
 errors include an estimate of the effect of varying \ptmin\ between 1.6 and
 2 GeV. An additional 7\% uncertainty is included in the single jet
 cross--section; this is the estimate for the contribution from soft processes.
 (Their contribution to the di--jet cross--section is negligible.) These
 numbers are preliminary; in particular, the error caused by fragmentation
 uncertainties has not yet been included. These numbers are reproduced equally
 well by the DG and LAC1,2 parametrizations; e.g., DG predicts 21.4 and 90.9
 pb for the first and second cross--section in eq.(\ref{topaz}), respectively.
 The predictions of the LAC3 parametrization are almost three times too large,
 so that this parametrization is clearly excluded; the DO+VMD parametrization
 is also disfavoured. The extraction of a partonic cross--section is an
 important step, since this allows to directly compare the predictions of
 theoretical models of \vqxqsq\ with their data; it also simplifies the
 comparison of experimental data from different groups.

 Finally, it should be mentioned that two LEP groups, ALEPH and DELPHI,
 have also presented first preliminary results om \gamgam\ scattering
 \cite{aleph,delphi}. The size of their data samples is only about 20\%
 of those of the TRISTAN groups; in addition, the very large annihilation
 cross section at $\rts \simeq m_Z$ poses special background problems
 \cite{dgolep}. Nevertheless, both groups confirm that traditional
 ``QPM+VMD'' models are in conflict with their data, while the inclusion
 of resolved photon contributions, as predicted from the DG parametrization,
 leads to satisfactory agreement between MC results and data. Notice that
 the good angular coverage of LEP detectors can give them an advantage
 when trying to find direct evidence for spectator jets, which so far have
 only been observed in \gamgam\ collisions via their contribution to the
 thrust distribution. Good angular coverage is also necessary for the
 study of events where one or both high$-p_T$ jets emerge at small angles,
 which would allow to extend the range of $x$ values probed by a given
 experiment.

 We have seen that the jet cross--sections and especially the resolved
 photon contributions fall off sharply with increasing $p_T$; e.g. in fig.
 \ref{tript}a the `2--res' processes dominate only upto $p_T
 \simeq 3$ GeV.  However, if we consider $d
 \sigma (jj) /d m_{jj} $ where $m_{jj}$ is the invariant
 mass of the two high$-p_T$ jets, then the resolved processes dominate to
 higher values. This can be understood from the fact that only 2--res processes
 receive contributions from diagrams where a gluon is exchanged in the
 $t-$ or $u-$ channel, leading to a more singular
 dependence  of the hard subprocess cross--section on the square of momentum
 transfer $\hat t$ for these contributions. Thus the
 study of invariant mass distributions, as well as rapidity
 distributions, can help us to get more detailed information on \qvph
 \cite{tristanac}. If one can measure the spectator jet energy
 instead of only using the spectator jet activity to `tag ' the `res'
 contributions, it may be possible to separate gluon initiated events
 from quark initiated ones. Since
 \glph\ is peaked at smaller values of $x$ as compared to $ \vec
 q_i^{\gamma} (x,Q^2), $
 the spectator jet energies will be higher for events initiated by
 gluons in the photon.

 As already stressed at the end of sec. 2, at present we have much less
 information about the gluon content of the photon than about its quark
 content. Unfortunately, the production of central jets with $p_T \geq 3$
 GeV is sensitive mainly to the region $x \geq 0.15$ at TRISTAN and LEP1
 energies\footnote{Recall that the \gamgam\ cms energy \wgg\ is usually
 much smaller than \rts, since $\fgme(x) \propto 1/x$.}. As a result, the
 study of high$-p_T$ jet production at these colliders is unlikely to
 discriminate between different ans\"atze for \gga, provided only that it
 is ``soft'' (this rules out LAC3, as we have seen). The production of
 open or hidden charm might offer better opportunities to get information
 about the gluon content of the photon. First of all, there is no need of
 a $p_T$ cut to get rid of a ``soft'' component, since there is none. The
 invariant energy going into the hard scattering can therefore be almost a
 factor of 2 smaller than for the clean sample of high$-p_T$ jets, and
 correspondingly smaller values of $x$ can be probed. Secondly, even if it
turns out that some $p_T$ cut has to be imposed to allow for the identification
 of charm events, the sensitivity to \gga\ is still greater than for inclusive
 jet production, because there is less background from resolved photon events
 initiated by quarks: The only 1--res contribution comes from $\gamma g$
 fusion, and the 2--res contribution from \qqbar\ annihilation is predicted to
 be very small. Any nonzero signal for \ccbar\ production via resolved photon
 processes would therefore allow a direct measurement of \gga.

 \begin{figure}
 \vspace {8cm}
 \caption{The total cross section for $\eplem \rightarrow \eplem \ccbar X$ as
 a function of \protect\rts.}
 \label{charmfig}
 \end{figure}

 In fig. \ref{charmfig} we compare predictions \cite{tristanac} for the total
 \ccbar\ cross section as calculated from the DG and LAC1 parametrizations,
 in the energy range covered by TRISTAN and LEP. For this and the following
 figures of this section, we have modified the estimate of the flux of
 resolved bremsstrahlung photons. The standard expression of ref. \cite{bkt}
 has been derived by integrating the photon propagator $\propto 1/P^2$
 over the full kinematically allowed range of the photon virtuality $-P^2$.
 However, if $P$ is larger than the scale $Q$ characterizing the hard
 scattering process, the picture of real (on--shell) partons residing ``in''
 the photon is no longer valid. Therefore the upper limit of the $P^2$
 integration should be of order $Q^2$. Moreover, it has been known for some
 time \cite{uematsu} that the parton content of virtual photons with
 $\Lambda_{\rm QCD}^2 < P^2 < Q^2$ is suppressed compared to the parton
 content of on--shell photons. We attempt a crude estimate of this
 effect by introducing a further suppression factor of 0.85 for the
 bremsstrahlung flux of resolved photons; this number has been estimated
 from numerical results of Rossi \cite{uematsu}\footnote{It has recently
 been pointed out \cite{borschu} that the suppression ought to be different
 for quarks and gluons. Since in perturbation theory gluons can only be
 radiated off quarks, which are themselves off--shell if $P^2 \neq 0$, their
 density in the photon drops faster with increasing virtuality of the photon
 than the quark content does.}. Altogether we thus have:
 \be \label{fbrems}
 f^{\rm brems}_{\gamma|e}(x) = 0.85 \frac {\alpha} {2 \pi}
 \left[ 1 + \left( 1-x \right)^2 \right] \ln \frac {Q^2} {m_e^2}.
 \ee
 The effect of this refinement will obviously be larger for larger ratio
 $s/Q^2$; this is why we did not introduce it in our predictions for jet
 production at TRISTAN energies and below.
 Of course, the formula of ref.\cite{bkt} is still applicable for the flux of
 bremsstrahlung photons interacting directly.

 We see that, with the exception
 of the immediate vicinity of the $Z$ pole, the two--photon cross section is
 {\em larger} than the one for the corresponding annihilation process
 $\eplem \rightarrow \ccbar$, by a factor of at least 8 (200) at \rts\ = 60
 (200) GeV. Secondly, just as in case of jet production, the contributions
 from resolved photon processes grow substantially faster with energy than that
 of the direct $\gamgam \rightarrow \ccbar$ process. However, at least for the
 more conservative DG parametrization, the direct contribution still dominates
 the total cross section even at \rts\ = 200 GeV. Cuts on the transverse
 momentum or angle of the produced charm quarks will further reduce the
 importance of the resolved photon contribution, since it has a softer
 $p_T$ spectrum and a more asymmetric angular distribution than the
 direct contribution. If the actual gluon content is indeed described by the
 DG parametrization, \ccbar\ production can therefore only be used to
 measure \gga\ if the direct contribution can be suppressed by detecting the
 spectator jet from the resolved photon, either directly or via the total
 thrust distribution. The GRV parametrization even predicts slightly (by 5 to
 10\%) smaller resolved photon contributions, due to the smaller value of
 \alpas\ that has to be used with this parametrization; note that at the
 rather low energy scales characteristic for charm production, \alpas\
 depends quite sensitively on $\Lambda_{\rm QCD}$.

 On the other hand, the LAC1 parametrization predicts the resolved photon
 contribution to dominate already for $\rts \geq 55$ GeV; at \rts\ = 200 GeV
 it predicts an almost 3 times larger total \ccbar\ cross section than the DG
parametrization does. (Similar results also hold for the LAC2 parametrization.)
 If the LAC1 predictions turn out to be close to the truth, isolation of the
 resolved photon contribution to \ccbar\ production should not be very
 difficult. However, even in this extreme case the 2--res contribution only
 amounts to 1.5\% of the total at the highest LEP energy; any resolved
 photon signal in \ccbar\ production can therefore safely be identified as
 stemming from 1--res photon--gluon fusion process.

 Another interesting process at existing \eplem\ colliders is the production
 of  $J/\Psi$ mesons in the reaction
 \be
 e^+ + e^- \rightarrow \gamma + \gamma \rightarrow J/\Psi + X.
 \ee
 In this case the relevant hard scattering process is $\gamma + g \rightarrow
 J/\psi + g$, which can be estimated from the colour singlet model
 \cite{berger}. Notice that this
 requires one of the photons to be resolved; in leading order in $\alpha$
 and \alpas, there is {\em no} direct contribution. Moreover, as in case of
 open charm production, the 2--res contribution is negligible.
 As a result the process is an extremely clean probe of \glph\ for
 $x \sim 0.01-0.05.$ The main problem in this case is the rather small
 total cross section; the DG parametrization predicts the cross section
 to grow from 0.5 pb at \rts\ = 60 GeV  to 4.0 pb at \rts\ = 200 GeV
 \cite{tristanac}. The LAC1,2 parametrizations again lead to more than 3
 times larger cross sections. However, almost certainly only the 12\% of
 all $J/\psi$ mesons that decay into \eplem\ or $\mu^+\mu^-$ pairs will
 be decetable. Moreover, there are indications from the photoproduction
 of $J/\psi$, which proceeds via the {\em same} hard scattering process,
 that the LO prediction of the colour singlet model might be too low by as
 much as a factor of 5 \cite{nmcpsi}. While this is good news as far as the
 observability of the $J/\psi$ signal at \eplem\ colliders is concerned,
 this large theoretical uncertainty means that at present this process
 cannot be used for a reliable measurement of the absolute size of the
 gluon component of the photon.\footnote{The large size of the ``k-factor''
 is not only due to the usual HO corrections, but presumably also includes
 corrections to the treatment of the $J/\psi$ meson as a non--relativistic
 \ccbar\ bound state. Already in a ``hybrid'' treatment, where the 1--loop
 corrected leptonic decay width of $J/\psi$ is used to determine the wave
 function at the origin which also determines the cross section, about 50\%
 higher rates are predicted.}

 Given a sufficiently large data sample, one might even attempt to look for
 \gamgam\ processes that occur only in higher orders in $\alpha$. An
 example is the study of the production of prompt photons in \gamgam\
 collisions via the processes,
 \bea
 \rm
 q^\gamma + \gamma & \rightarrow & {\rm q } + \gamma \nonumber \\
 \rm q^\gamma +  \bar q^\gamma   &\rightarrow &\rm {\gamma + g  }\nonumber\\
 \rm g^\gamma + q^\gamma & \rightarrow & q + \gamma
 \eea
 These processes, though suppressed by  a factor of $\alpha $ compared to
 the case of jet production, have the advantage of having a cleaner
 final state. In the PEP to TRISTAN energy range the 1--res contribution
 clearly dominates \cite{dg2}; requiring $p_T^{\gamma} (= p_T^{\rm jet}$
 in leading order) to be larger than 1.5 GeV (a value very close to
 the value of \ptmin\ as determined from the AMY jet analysis \cite{amy}
 described above), the DG parametrization predicts a total $\gamma + $jet
 cross section of about 1 pb at \rts\ = 60 GeV. At LEP energies the
 2--res contributions become more significant, and might allow to extract
 additional nontrivial information about the parton content of the
 photon \cite{dg2}.

 We have seen repeatedly that, for fixed transverse momenta of the particles
 produced in the hard scattering process or fixed invariant mass of the
 system produced in that scattering, the importance of resolved photon
 events increases quite rapidly with increasing beam energy; under the
 same circumstances, the ratio of \gamgam\ to annihilation events also
 grows rapidly with \rts, as we have seen for the example of the total charm
 production cross--section. At future \eplem\ colliders it might therefore
 no longer be possible to consider \gamgam\ events as a background that
 can be suppressed easily.

 This is demonstrated in fig. \ref{linacpt},
 \begin{figure}[hbt]
 \vspace {8 cm}
 \caption{Cross-section for \gamgam\ $\rightarrow \rm jets $ as a
 function of $\rm \protect\sqrt s$  for \ptmin\ = 5 GeV. }
 \label{linacpt}
 \end{figure}
 where we show the energy dependence of the total cross--section for the
 production of a pair of central jets (with rapidity $|y_{1,2}| \leq 2$)
 with $p_T \geq 5$ GeV in \gamgam\ collisions at high energy \eplem\
 colliders. Here we have conservatively ignored all effects of
 beamstrahlung (see below), and have used eq.(\ref{fbrems}) for the
 flux of resolved photons. Nevertheless, even the DG parametrization
 predicts resolved photon contributions to be dominant already at
 \rts\ = 200 GeV; at \rts\ = 500 GeV, which is now foreseen as the
 likely operating energy of the next (linear) \eplem\ collider, resolved
 photon processes are predicted to dominate the direct one by a ratio of
 6:1. Once again the LAC2 parametrization predicts both a more rapid
 increase with energy, and a considerably larger absolute value, of
 the $\gamgam \rightarrow$ jets cross--section. Notice finally that
 the annihilation cross section at \rts\ = 500 GeV only amounts to
 0.4 pb for $\mu^+ \mu^-$ pairs, and 8 pb for $W^+ W^-$ pairs; this is
 to be compared to a \gamgam\ cross--section of at least 150 pb at the
 same energy, even using the relatively strong cuts of fig.
 \ref{linacpt}; this cross--section is as large as 500 pb if the photon
 structure is better described by the LAC1 parametrization.

 In fig. \ref{linacpt} we have only included the bremsstrahlung contribution
 to the photon spectrum. However, it is well known that synchrotron radiation
 makes the construction of \eplem\ storage rings with \rts\ significantly
 beyond the reach of LEP2 prohibitively expensive; one will therefore have to
 use linear colliders (linacs) if higher energies are to be reached in
 \eplem\ collisions. In such linacs, each electron or positron bunch has only
 a single chance to produce a reaction; at the same time, the total
 luminosity of the machine has to grow $\propto s$ if a useful rate of
 annihilation events is to be maintained. These two constraints imply that
 the luminosity per bunch crossing has to be much larger than at present
 storage rings; this in turn necessitates the use of very dense bunches,
 with transverse dimensions of the order of a few dozen nm. This leads to
 a large charge density, which produces very strong electromagnetic
 fields. Immediately before and during bunch collisions the particles in
 one bunch feel the field produced by the other bunch, and are accelerated.
 The radiation produced by this acceleration is known as beamstrahlung
 \cite{beam}.

 This qualitative discussion shows that machines with large luminosity per
 bunch crossing generally produce more beamstrahlung\footnote{Beamstrahlung
 can be reduced while keeping the luminosity fixed if elliptic bunches with
 large aspect ratio (the ratio between the semi--major and semi--minor axis)
 are employed \cite{blankdrell}. However, the steering of very flat beams
 is expected to be even more difficult than for round beams, which puts an
 upper limit on the aspect ratio. Typical aspect ratios for present designs
 range from a few dozen to about 100.}. The necessary luminosity per bunch
 crossing is obviously inversely proportional to the number of bunch
 collisions per second; this number in turn depends on the design of the
 accelerating structures.

 Some typical examples of beamstrahlung spectra for \eplem\ colliders
 operating at \rts\ = 500 GeV are shown in fig. \ref{beam} \cite{fatdg};
 these curves have been computed using approximate analytical expressions given
 in ref.\cite{chen}. The acronyms P--G, P--F, D--D and T stand for the
 Palmer--G, Palmer--F, DESY--Darmstadt and TESLA designs, respectively,
 while wbb (nbb) denotes the wide (narrow) band beam option of the D--D
 design. We see that machines that utilize accelerating RF fields with
 wavelength in the ``X band'' region (P--G, P--F as well as the Japanese
 Linear Collider JLC, whose first stage is somewhat similar to the
 Palmer--F design) have harder beamstrahlung spectra than designs using the
 longer wavelengths of the ``S band'' (D--D), or the TESLA design, which
 is based on superconducting cavities. For comparison we also show the
 Weizs\"acker--Williams (WW) bremsstrahlung spectrum (dotted curve). For
 large fractional photon momentum $x$, the beamstrahlung contribution is
 exponentially suppressed; this end of the spectrum is therefore still
 dominated by the bremsstrahlung contribution. However, at smaller values
 of $x$ beamstrahlung photons are more abundant; the cross--over point
 between the regions dominated by beam-- and bremsstrahlung depends
 sensitively on the machine parameters.

 \begin{figure}[hbt]
 \vspace {8 cm}
 \caption{The beamstrahlung photon spectrum of 4 typical designs of
 500 GeV \protect\eplem linacs, as well as of bremsstrahlung photons
 (dotted) and of backscattered laser photons (dot--dashed). From
 \protect\cite{fatdg}.}
 \label{beam}
 \end{figure}

 Finally, it has been pointed out \cite{telnov} that an \eplem\ collider
 can be converted into a \gamgam\ collider by shining very intense laser
 light on the particle beams; some laser photons then undergo Compton
 backscattering. The dash--dotted curve in fig. \ref{beam} shows the
 spectrum that results if the laser energy is chosen such that the
 invariant mass of a laser and a backscattered photon is just below
 2 $m_e$, and both laser and electron beam are unpolarized. Notice that
 these backscattered photons, as well as beamstrahlung photons, are
 truly on--shell, unlike bremsstrahlung photons.

 Obviously beamstrahlung can greatly enhance rates of two--photon events.
 For example, had we included the beamstrahlung contribution to \fgme\
 in fig. \ref{linacpt}, the cross--section would have grown \cite{fatdg} to
 between 180 pb (for TESLA) and 4.5 nb (for Palmer--G). Since the luminosity of
 those designs is 2 (6) $\cdot 10^{33} {\rm cm}^{-2} {\rm sec}^{-1}$, this
 corresponds to approximately 4 (250) {\em million} events with total
 hard $E_T >$ 10 GeV per year for the TESLA (Palmer--G) collider! Of course,
 in principle one can get rid of most of these events by setting a rather
 high trigger threshold on the transverse momentum of the jets, or the
 total $E_T$ in the event\footnote{A cut on the visible energy $E_{vis}$
 would be less effective; \gamgam\ events often have $E_{vis} \gg E_T$,
 since jets are often produced at small angles, and since spectator jets
 contribute to $E_{vis}$ but only little to $E_T$.}. However, then one
 risks to lose interesting annihilation events containing massive stable
 neutral particles, as predicted e.g. by supersymmetric theories. Moreover,
 the \gamgam\ events are interesting in their own right. In our view it
 is therefore preferable to use a low trigger threshold, even if this means
 that the amount of data to be handled is rather large for \eplem\
 colliders; it is still small compared to the amount of information that
 has to be manipulated at typical LHC or SSC detectors.

 Beamstrahlung also changes the electron spectrum \cite{chen}; obviously
 an electron will lose some of its energy when emitting a hard photon.
 This effect has to be added to the smearing of the beam energy due to
 the machine parameter independent initial state radiation. For designs
 with hard beamstrahlung spectrum (e.g., Palmer--G), the \eplem\ luminosity
 spectrum is distorted by beamstrahlung even for energies far below the
 nominal \rts\ of the collider. At small invariant masses one thus generally
 has a competition between \gamgam\ and \eplem\ events. This is demonstrated in
 fig. \ref{mjj}, where we show the invariant mass spectrum of events with
 two central jets with $p_T \geq 20$ GeV \cite{fatdg}.
 \begin{figure}[hbt]
 \vspace {8 cm}
 \caption{The invariant mass distribution of events with two central jets
 with $p_T \geq 20$ GeV. The resolved \protect\gamgam\ contributions have been
 estimated \protect\cite{fatdg} using the DG parametrization.}
 \label{mjj}
 \end{figure}
 The annihilation contribution
 exhibits a prominent peak at \rts\ = $M_Z$; by comparing events in that
 peak with events with $M_{jj} \simeq \rts$, one can hope to study the
 QCD evolution of the hadronic sytem with increasing invariant mass in a
 single detector, thereby reducing experimental (systematic) errors. However,
 this figure shows that at this collider it would be very difficult to
 extract a clean sample of annihilation events with $M_{jj} \simeq M_Z$;
 in spite of the rather severe cut on $p_T$ which reduces the \gamgam\
 contribution considerably, the $Z$ peak will hardly stand out in the
 total sample of di--jet events once detector resolution effects are
 included. In colliders with soft beamstrahlung spectrum (TESLA or
 the nbb option of D--D) the annihilation cross--section at $M_{jj} = M_Z$
 is reduced by a factor of 3, but the \gamgam\ contribution is almost
 30 times smaller than at Palmer--G, enabling one to isolate a rather
 clean sample of annihilation events from the $Z$ peak.

 As a final example of the importance of beamstrahlung we list in table 1
 estimates of total \ccbar\ and $J/\psi$ production cross sections, using the
 DG parametrization. Since these processes are sensitive to the region of small
 $x$, even the soft beamstrahlung spectrum of the TESLA collider leads to
 a sizable enhancement of the rate. This is especially true for the direct
 contribution, whose cross--section decreases with increasing \wgg, unlike
 those for the resolved photon contributions; without beamstrahlung the direct
 and 1--res total \ccbar\ cross sections would only amount to 0.6 and 0.85 nb,
 respectively. Due to the different dependence on \wgg, resolved photon
 events are more important at colliders with harder beamstrahlung spectrum.
 However, even for the Palmer--G design we find \cite{fatdg} that the
 cross--section for the production of central \ccbar\ pairs with $p_T > 5$
 GeV is dominated by the direct contribution. This is because the 1--res
 contribution has a softer $p_T$ spectrum and, due to the asymmetric
 initial state, is peaked at small angles. If, on the other hand, the \eplem\
 collider is converted into a \gamgam\ collider, even the 2--res contribution
 will be larger than the direct one; notice that in this case the 2--photon
 luminosity actually falls at small \wgg.

 \vspace*{6mm}
 \noindent
 {\bf Table 1:} Total \ccbar\ cross--sections from two--photon processes at the
 4 \eplem\ colliders of fig.\ref{beam}, as well as for a \gamgam\ collider made
 from an \eplem\ collider with \rts\ = 500 GeV.
 We have used the DG parametrization to
 estimate the resolved photon contributions. $\sigma( \qqbar)$ and $\sigma(gg)$
 stand for the 2--res \qqbar\ annihilation and gluon fusion cross--sections,
 $\sigma(\gamma g)$ for the 1--res photon gluon fusion contribution, and
 $\sigma(\gamgam)$ for the direct contribution; $\sigma(J/\psi)$ is the 1--res
 $\gamma + g \rightarrow J/\psi + g$ cross--section in the color singlet model.
 All cross--sections are in nb.
 \begin{center}
 \begin{tabular}{|c||c|c|c|c|c|c|}
 \hline
 Collider & $\sigma(\qqbar)$ & $\sigma(gg)$ & $\sigma(\gamma g)$ &
 $\sigma(\gamgam)$ & $\sigma($tot) & $\sigma(J/\psi)$ \\
 \hline
 T & 0.010 & 0.038 & 1.8 & 2.2 & 4.0 & 0.014 \\
 D--D(wbb) & 0.041 & 0.11 & 7.0 & 6.4 & 13.5 & 0.053 \\
 P--F & 0.017 & 0.08 & 4.0 & 2.4 & 6.4 & 0.030 \\
 P--G & 0.14 & 1.1 & 38 & 9.9 & 49 & 0.28 \\
 \gamgam(500) & 0.13 & 7.6 & 130 & 0.14 & 140 & 0.89 \\
 \hline
 \end{tabular}
 \end{center}
 \vspace*{6mm}

 Qualitatively similar results hold for total $b \bar{b}$ production,
 except that the cross--sections are smaller by a factor between 100 and
 200. It has been claimed \cite{halzentop} that at future linacs total
 \ttbar\ production might also be dominated by the \gamgam\ contribution.
 We find \cite{fatdg}, however, that even at the Palmer--G collider the
 \gamgam\  contribution amounts to at most 5\% of the total; for the other
 designs of 500 GeV linacs this number is closer to 1\%. Even for the third
 stage of the JLC, which operates at \rts\ = 1.5 TeV and also has a rather hard
 beamstrahlung spectrum, the annihilation contribution is still dominant if
 $m_t > 130$ GeV. At such very high energy colliders, beamstrahlung and
 initial state radiation also increase the annihilation contribution by as much
 as 60\%, due to the reduction of the average centre--of--mass energy of
 \eplem\ pairs. In principle one could increase beamstrahlung even further,
 e.g. by using round beams. However, we will argue in sec. 5 that in this case
one will have to deal with qualitatively new beamstrahlung induced backgrounds,
 including the existence of an ``underlying event'' which will make experiments
 at such \eplem\ linacs similar to those at hadron colliders, so that the
 detailed study of \ttbar\ events will become very difficult. In contrast, most
 of the \eplem\ colliders discussed here could quite easily accumulate a
 clean sample of \ttbar\ events from \eplem\ annihilation. We therefore see
 no advantage of operating future linacs in the domain of high beamstrahlung.

 Of course, at present predictions for total \ccbar\ and $b \bar b$ production
 cross sections at high energy linacs suffer from large uncertainties,
 since one is probing the parton content of the photon at values of $x$ as
 small as $10^{-3}$, where no experimental information exists so far. With the
 advent of the $ep$ collider HERA this is expected to change soon, however, as
 we discuss in the next section.

 \renewcommand{\theequation}{4.\arabic{equation}}
 \setcounter{equation}{0}
 \setcounter{footnote}{0}
 \section*{4) Resolved Photon Reactions in $\gamma p$ Scattering}
 The discussion in earlier sections indicates that we need to probe
 the parton content of the photon, especially the gluon content, at small
 values of
 $x$ and large \qsq. For reasons discussed earlier, DIS experiments are
 limited by statistics in the region of large \qsq\ and probe
 $G^\gamma $ only indirectly.  Jet production in \gamgam\ collisions at
 TRISTAN and LEP will certainly provide useful information. But the only other
 possibility to go to higher values of \rts\ for photon interactions
 and hence increase the range of $x$ and \qsq\ values at which the
 photon can be probed, is at present the high energy $ep$ collider HERA.

 The suggestion to use $\gamma p $ collisions to study \qvph\ is
 not new \cite{owens79}. Theoretically the situation is actually somewhat
 simpler than for real \gamgam\ scattering, since we only have to deal with
 two classes of contributions: Direct ones, where the photon directly interacts
 with the partons in the photon; and resolved photon reactions, where
 the partons in the photon scatter off partons in the proton.
 The low energy of photon beams available at fixed target experiments reduces
 the contribution of `res' processes in current experiments but it
 still plays an important role \cite{owens79,do,baau}. At the high
 energy HERA collider with an $ep$ center of mass energy $\sim 300$
 GeV, the situation is quite different. In a large number of QCD processes
 such as high $p_T$ jet production \cite{dg3,st,baer}, heavy flavour
 production \cite{dg3,elku}, direct photon
 production \cite{bawa1} and Drell Yan lepton pair
 production \cite{kust1} the hadronic structure of the photon not
 only plays an extremely important role but even dominates in some cases.
 The cross--section for the various QCD processes is given by
 expressions very similar to eq.(\ref{csec}), where one replaces one
 photon by the proton and includes the partonic subprocesses corresponding
 to the QCD process under consideration. We now describe some reactions in
 more detail.

 \subsection*{4a) $ep \rightarrow$ jets $+ \ X$}
 We start with a discussion of inclusive jet pair production, which offers the
 highest cross--section of all hard scattering processes at HERA. Here the
 direct processes are the same as the 1--res processes of \gamgam\ scattering,
 and the resolved photon reactions correspond to the 2--res contributions
 to \gamgam\ collisions.

 In fig. \ref{rsigma} we show the ratio $R_{\sigma}$ of the cross--sections
 of the resolved and direct processes as a function of the transverse
 momentum $p_T$ of the jets \cite{dg3}.
 \begin{figure}
 \vspace{8 cm}
 \caption{Ratio of resolved and direct contributions for
 $d \sigma (ep \rightarrow$ jets$)/  d p_T$ as a function of $p_T$
 \protect\cite{dg3}.}
 \label{rsigma}
 \end{figure}
 We see that even the DG parametrization predicts the resolved photon
 conbtribution to be larger than the direct one out to $p_T \simeq 35$
 GeV. In $\gamma p$ collisions we have to pay the price (in terms of
 reduced phase space) of producing an additional spectator jet only once
 when we want to gain access to the QCD \totwo\ scattering processes, whose
 matrix elements are enhanced by gluon exchange in the $t-$ or
 $u-$channel, as discussed earlier; in \gamgam\ collisions this price has
 to be paid twice, since both photons have to be resolved. Moreover, the
 proton has a relatively larger gluonic component than the photon, at least
 if the DG parametrization is close to the truth. This further favours
 resolved photon processes over direct ones, since the QCD \totwo\
 matrix elements containing gluons in the initial and final state are
 enhanced by colour factors, while the direct $\gamma g$ fusion process is
 actually colour--suppressed (by a factor of 3/8) compared to $\gamma q$
 scattering. These two observations explain why at HERA resolved photon
 processes dominate jet production out to larger values of $x_T \equiv
 2 p_T/\rts$ than at \eplem\ colliders; see fig.\ref{tript}.
 Fig.\ref{rsigma} also illustrates that existing
 parametrizations of proton structure functions (DO2 \cite{dop} vs.
 GHR \cite{ghr}) differ much less than those for the photon, at least in
 the region $x \geq 0.01$ relevant for the production of high$-p_T$ jets.

 In fig.\ref{herapt} a,b we show the $p_T$ spectrum in absolute units.
 In these figures, we also show the contributions from different final
 states separately; for most events the parton composition of the initial
 and final states are identical. We see that, unlike at hadron colliders,
 the $gg$ final state dominates only at very low values of $p_T$, below
 3 GeV for the DG parametrization. This is because the difference in shape
 between quark and gluon distribution functions is larger for photons
 than for nucleons. In the latter case all parton densities fall with
 increasing $x$, while $x \cdot q^{\gamma}$ has a maximum at $x \simeq 0.9$,
 as shown in fig.\ref{parcomp}.

 \begin{figure}
 \vspace{8 cm}
 \caption{Resolved (a) and direct (b) contribution to two--jet production
 at HERA, where the different final states are shown separately;
 $q$ denotes a quark or anti--quark of any flavour. We have used the
 DO2 and DG parametrizations for the proton and the photon, respectively.
 From ref.\protect\cite{dg3}.}
 \label{herapt}
 \end{figure}

 For most of the $p_T$ range where resolved photon contributions
 dominate, the largest contribution comes from the mixed $qg$ final
 state. The DG parametrization predicts that in most cases the quark comes
 from the photon and the gluon from the proton; this is again a result
 of the relative softness of \gga\ and the large gluon content of nucleons.
 Finally, fig. \ref{herapt}b shows that the direct contribution is only
 dominated by photon--gluon fusion for values of $p_T$ where the total
 di--jet cross--section is dominated by res contributions; this might make
 it difficult to extract the gluon density of the nucleon from measurements
 of jet production at HERA.

 Of course, in principle direct and res events can be distinguished by the
 presence of the spectator jet from the photon going in the electron beam
 direction, which is the hallmark signature of resolved photons. However,
 while there are arguments suggesting \cite{dg3} that this jet should be
 rather broad and hence easily detectable in most cases, the exact value of
 the efficiency for tagging on this jet clearly depends on the details of
 the jet formation model (i.e., it is not an ``infrared safe quantity''),
 as well as on the detector acceptances. It is therefore tempting to try and
 find differences between direct and res contributions in the distributions
 of the high$-p_T$ jets themselves, which can be predicted directly from
 perturbative QCD.

 One possibility \cite{baer} is to look at the cross--section as a function
 of the centre--of--mass scattering angle. Due to the presence
 of diagrams with gluon exchange in the $t-$ or $u-$channel the resolved
 photon contribution will be more strongly peaked at small angles than the
 direct contribution.

 Another possibility \cite{dg3} is to study the triple--differential
 cross--section $d \sigma / d p_T d y_1 d y_2$, where the $y_i$ are
 the rapidities of the two high$-p_T$ jets. The results of fig.\ref{herapt}
 show that the cross--section should be large enough to allow such detailed
 studies even with less than the full HERA design luminosity of about
 100 pb$^{-1}$/yr. In this case we can make use of purely kinematical
 considerations to separate the two classes of contributions. Obviously a
 parton ``in'' a photon will have less energy than the photon itself. For
 a given invariant mass of the produced partonic system, the parton from
 the proton will therefore have to supply more energy in resolved photon
 events than in direct ones; this results in a boost of the partonic
 system in the direction of the proton beam. Since, as we have
 emphasized repeatedly, \gga\ is expected to be much softer than the
 $q_i^{\gamma}$\footnote{It is worth mentioning that the validity of this
 statement rests on the fact that analyses of TRISTAN data have already
 ruled out the LAC3 parametrization, as discussed in the previous
 section.}, this boost will be stronger if the parton that is ``pulled out''
 of the photon is a gluon. Parametrizations with larger \gga\ will thus
 tend to predict a rapidity distribution that is more strongly peaked at
 larger rapidities.

 \begin{figure}
 \vspace{8 cm}
 \caption{Comparison \protect\cite{dg3} of the shape of the rapidity
 distribution of jets produced at HERA. Note that the dotted and
 dashed curves have been normalized, as described in the text.}
 \label{herarap}
 \end{figure}

 This is demonstrated in fig.\ref{herarap}, which shows the shape of
 the rapidity distributions (for $y_1 = y_2 \equiv y$) at $p_T = 10$
 GeV. In order to avoid ``k-factor'' uncertainties, all curves have been
 normalized to give the same single--differential cross--section
 $d \sigma / d p_T = 8.8$
 nb/GeV at $p_T = 10$ GeV. The DO+VMD distribution with its rather large
 gluon content leads to a much more pronounced peak at $y \simeq 2.2$
 than the DG parametrization does, although the position of the peak is
 not shifted very much; note that $y \rightarrow y_{\rm max}$ corresponds
 to $x_p \rightarrow 1$, where $\vec{q}^p$ vanishes. On the other hand,
 a toy model with zero gluon content predicts the peak in the rapidity
 distribution to be shifted towards smaller $y$ by about 1.5 units.
 We have already seen that AMY data require \cite{amy} a nonzero \gga,
 but they cannot distinguish between the DG and DO+VMD parametrizations;
 from fig.\ref{herarap} we can conclude that jet studies at HERA should
 allow much more detailed analyses, due to the large event rates even at
 rather large values of $p_T$, where the choice of \ptmin\ becomes
 irrelevant and the analysis is less sensitive to the details of the
 jet fragmentation scheme. Finally, it should be mentioned that the
 direct contribution actually peaks at the smallest possible value of $y$.
 Here $x_p$ approaches its kinematical minimum ($= x_T^2$), while nearly
 the whole electron energy has to be transmitted to the photon; this is
 quite unlikely, but the rapid increase of $\vec{q}^p$ at small $x_p$
 over--compensates this suppression factor.

 We thus see that the presence of resolved photon contributions leads to
 3 qualitative predictions \cite{dg3} that ought to be testable
 quite easily: Large jet production cross sections; spectator jets from
 the photon; and a rapidity distribution that is peaked at positive $y$
 (corresponding to the proton beam direction). Even though HERA experiments
 so far have only taken a few nb$^{-1}$ of data, they are already starting
 to confirm these predictions.

 In fig.\ref{zeuspt} we show the transverse energy spectrum of the
 photoproduction events identified by the ZEUS collaboration \cite{zeus} in
 the first (pilot) run of HERA.
 \begin{figure}
 \vspace{8 cm}
 \caption{The total $ep$ cross--section measured \protect\cite{zeus}
 for transverse energies larger than $E_T^0$. The curve is the HERWIG
 prediction, using the DG parametrization with \ptmin\ = 1.5 GeV.}
 \label{zeuspt}
 \end{figure}
 For $E_T > 10$ GeV the soft (VMD)
 contribution is found to be negligible; the events in this region therefore
 have to be explained by hard scattering processes. The dashed curve shows
 the prediction of the HERWIG generator \cite{herwig} using the DG
 parametrization with \ptmin\ = 1.5 GeV; we see that it describes the data
 quite well. Recall that almost the same value of \ptmin\ has been found
 to describe the TRISTAN data. From their data sample ZEUS extracts a
 total cross--section for the production of events with $E_T > 10$ GeV
 of $2.4 \pm 0.1 \pm 0.7 \ \mu$b. A glance at fig.\ref{herapt} shows
 immediately that a cross--section of this size cannot be explained from
 direct processes alone; indeed, the detailed MC study of the ZEUS group
 shows that without the resolved photon contributions theory falls short of the
 data by at least one order of magnitude. Similar results have been
 reported by the H1 collaboration \cite{h1}.

 Both groups also find evidence for the spectator jet from the photon in
 their data. As an example we show in fig.\ref{enflow} the energy flow
 measured \cite{h1} in the H1 calorimeter as a function of the angle
 ($\theta = 0$ is the direction of the proton beam); only events where
 both high$-p_T$ jets emerge at $\theta \leq 100^{\circ}$ have been included.
 \begin{figure}
 \vspace{8 cm}
 \caption{Histogram of energy flow per event versus polar angle. The
 open points represent the data, while the full and dotted lines give the
 MC prediction with and without resolved photon contributions. From
 ref.\protect\cite{h1}.}
 \label{enflow}
 \end{figure}
 At small angles a large amount of energy is deposited by the proton
 remnants. At intermediate angles both soft and hard processes contribute.
 However, direct events are unable to populate the region around the
 electron beam direction, in conflict with the data, which show a constant
 or even slowly rising energy deposition at $\theta \simeq 180^{\circ}$.
 This is well described by the MC generator once resolved photon contributions
 are included. ZEUS also finds \cite{zeus} evidence for the spectator jet
 in their sample of reconstructed jet events: Some events have sizable
 energy deposit around the electron beam direction even though all high$-p_T$
 jets are 2 or more units of rapidity away. Their data indicate that the
 efficiency for tagging on this spectator jet should be around 40\%, in
 qualitative agreement with earlier MC studies \cite{dag}.

 Finally, H1 also finds \cite{h1} that their jets populate a quite different
 angular (or rapidity) region than what one would expect from direct events.
 In particular, they find that a large fraction of their events have one
 or both jets relatively close to the proton beam direction; we have seen
 in fig.\ref{herarap} that direct $\gamma p$ scattering produces jets
 preferably at large angles. Our three main predictions are therefore
 all borne out at least qualitatively by the data; we are looking forward
 to more detailed analyses of higher statistics data samples.

 Before closing this subsection we should point out that recently first
 results of next--to--leading order calculations of the photoproduction of
 jets at HERA energies have become available. NLO corrections to the
 direct processes have been computed already some time ago by Aurenche
 et al. and independently by Baer et al. \cite{baau}; they have recently been
 re--done and applied to HERA in ref.\cite{boedeker}. The production of three
 jets in direct and resolved photon interactions has been studied in
 ref.\cite{baer}. Full NLO corrections to the resolved photon contribution
 have been computed in ref.\cite{salesch}. Finally, so far the only paper
 that includes a full next--to--leading order calculation of all contributions
 to jet production
 is ref.\cite{gorstor}. Such a comprehensive analysis is necessary since
 some divergent corrections to the direct process have to be absorbed in
 the parton distribution functions of the photon, thereby blurring the
 distinction between direct and resolved photon contributions. These
 studies indicate that inclusion of NLO corrections reduces the artificial
 dependence of the cross--section on factorization and renormalization
 scales. However, if these scales are chosen to be equal to the transverse
 momentum of the jets and a cone size $\Delta R = 0.7$ is chosen in the jet
 definitions, NLO corrections appear to be quite modest. The results of
 this subsection, which have been obtained from leading order calculations,
 should therefore retain their validity also in NLO.

 \subsection*{4b) Heavy Quark Production}
 We have seen in the last subsection that inclusive jet production at HERA
 will probably only allow a rather indirect determination of \gga\, due to
 the large background from quark--initiated resolved photon events. Just as
 in case of \gamgam\ scattering, one can enhance the importance of
 gluon--initiated processes by studying specific final states. Among those,
 the production of a pair of heavy quarks offers the largest cross--section.
 We focus here on $b$--quarks, which should be easier to identify than
 $c$ quarks, and where fragmentation effects should be smaller.

 The total \bbbar\ cross--section at HERA as predicted from the DG
 parametrization is \cite{dg3,elku} about 1 nb, which corresponds to
 100,000 \bbbar\ pairs per year. Unfortunately the resolved photon
 contribution only amounts to about 20\% of the total; a separation of the
 two classes of contributions thus becomes mandatory if \bbbar\ production is
 to be used to determine \gga. Fortunately we have seen at the end of the
 previous subsection that it seems to be possible to tag spectator jets from
 the photon with reasonable efficiency; this should allow to accumulate a
 rather clean sample of resolved \bbbar\ events.

 In fig.\ref{bpt} we show the $p_T$ spectrum of the $b$ (or $\bar b$) quark
 for the symmetric configuration $y_1 = y_2 \equiv y$; the res contribution
 has again been estimated from the DG parametrization. We see that the slope
 of the spectrum at high $p_T$ depends quite sensitively on $y$; recall that
 large $y$ correspond to large $x_p$, where the parton densities in the
 nucleon decrease rapidly. One can also conclude that a detailed study of
 the resolved photon contribution will only be possible if $b$ quarks with
 $p_T$ below 10 GeV can be identified efficiently and reliably; otherwise the
 rate will be too small.

 \begin{figure}
 \vspace{8 cm}
 \caption{$p_T$ distribution of the $b$ (or $\bar b$) quark produced in
 photoproduction events at HERA as predicted \protect\cite{dg3} from the DG
 parametrization, for $y_1 = y_2 \equiv y$.}
 \label{bpt}
 \end{figure}

 Fig.\ref{bpt} also shows that resolved photon contributions are much more
 important at large $y$. After the discussion of the rapidity distribution of
 jets in the previous subsection this should not be surprising; since the
 res contribution to \bbbar\ pair production is dominated by $gg$ fusion,
 res \bbbar\ events will usually undergo a strong boost in the proton beam
 direction. This is further illustrated in fig.\ref{brap}, where we show the
 \bbbar\ cross section as a function of $y_2$ for fixed $y_1$, at $p_T = 5$
 GeV. Notice that in this figure the res contribution has been computed from
 the DO+VMD parametrization, which predicts a ratio of direct to resolved
 contributions of about 2:1, as opposed to 4:1 for the DG parametrization.
 \begin{figure}
 \vspace{8 cm}
 \caption{Rapidity distribution of the $b$ (or $\bar b$) quark produced in
 photoproduction events at HERA as predicted \protect\cite{dg3} from the
 DO+VMD parametrization, for fixed value of the other rapidity and $p_T = 5$
 GeV.}
 \label{brap}
 \end{figure}
 We have seen above that the efficiency for tagging the photonic spectator
 jet might be around 40 to 50\% for generic high$-p_T$ jet events; it might
 be somewhat smaller for the more spherical \bbbar\ events. Simply requiring
 the absence of such a tag would thus leave a \bbbar\ sample that still
 contains 20 to 30\% resolved photon events, if \gga\ is similar to the
 DO+VMD parametrization; this could complicate the extraction of the gluon
 content of the nucleon from studies of \bbbar\ pair production at HERA.
 Fortunately fig.\ref{brap} shows that even the large res contribution
 predicted by the DO+VMD parametrization can be suppressed to an
 insignificant level by requiring either the $b$ {\em or} the $\bar b$
 to emerge at small (negative) rapidity; this restriction still allows to
 probe $G^p(x_p)$ for $x_p$ between $2 \cdot 10^{-3}$ and 1 by varying
 $y_2$ within its kinematically allowed limits. The same conclusion holds
 for all parametrizations of \vqxqsq\ that predict $\gga(x)$ to be
 soft, as seems to be required by TRISTAN data. By separating the total
 \bbbar\ sample into events with and without a photonic spectator jet, and
 studying the rapidity distribution in each sample, it should therefore be
 possible to extract important information about both $G^p$ and \gga\
 from \bbbar\ pair production at HERA.

 Finally, we remark that so far only partial NLO calculations for the
 photoproduction of heavy quark pairs exist \cite{elku,smith,donew}. In these
 papers the corrections to the direct process are included, but the
 resolved photon contribution, which occurs at the same order in \alpas,
 has only been included at tree level.

 \subsection*{4c) Direct Photon Production}
 Another process that can be studied at HERA is the production of hard
 direct photons in $e p \rightarrow e \gamma X$ \cite{oldphot}. Of course,
 the cross--section is now ${\cal O}(\alpha^3)$ and thus approximately two
 orders of magnitude smaller than the jet cross--section. On the other hand,
the direction and energy of a hard photon can be determined much more precisely
 than those of a jet; this should help in the reconstruction of the
 Bjorken--x variables of the partons participating in the hard scattering.

 In ref.\cite{bawa1} a fairly comprehensive study of this reaction has been
 presented for HERA energies. NLO corrections to the direct process
 $\gamma q \rightarrow \gamma q$ \cite{maria} are included, but the resolved
 photon contributions ($gq \rightarrow \gamma q$ and $q \bar{q} \rightarrow
 \gamma g$) are treated at the Born level. If a hard photon within a jet
 can be detected, one can also study the fragmentation of a parton into a
 photon, which is the inverse of $\gamma \rightarrow$ parton splitting
 described by the parton densities in the photon. Even if these
 contributions are included, resolved photon processes dominate the total
 cross--section only for $p_T^{\gamma} \leq 15$ GeV, according to the DG
 parametrization. The reason is that again one has to produce two
 additional jets (the spectator jet from the photon, and the remnants of
 parton $\rightarrow$ photon fragmentation) before the QCD
 \totwo\ processes become accessible.

 Nevertheless the study of photons with $p_T \simeq 5$ GeV or so should yield
 information about \qvph, especially \gga. Kinematics again implies that
 events of the type $q^p g^{\gamma} \rightarrow q \gamma$ should be strongly
 boosted in the proton direction; in addition, the hard matrix element favours
 the photon to emerge close to the direction of the incident quark. The
 combination of these two effects implies \cite{bawa2} that res contributions
 dominate at small angles relative to the proton beam direction, as shown in
 fig.\ref{gamspec}.
 \begin{figure}
 \vspace{10 cm}
 \caption{Energy spectrum of photons produced in $\gamma p$ scattering at
 $s = 30,000 \rm{GeV}^2$, as predicted \protect\cite{bawa2} from the DG
 parametrization.}
 \label{gamspec}
 \end{figure}
 Here a fixed energy of the incoming photon has been
 assumed, $E_{\gamma} = 9$ GeV; experimentally this means that the ougoing
 electron has to be tagged in a forward spectrometer, which is also used for
 luminosity measurements. Since the transverse momentum of the outgoing photon
 has been fixed to 5 GeV, there is a one--to--one relation between the energy
 and the angle of the photon, with small angles corresponding to large
 energies. The coverage of the electromagnetic calorimeter of HERA experiments
 starts approximately 4 degrees from the proton beam pipe \cite{bawa2}; this
 means that photons with energy as large as 110 GeV should be detected with
 sizable rates. Notice also that at this angle even the DG parametrization
 predicts the resolved photon contribution to be a full order of magnitude
 above the direct one. Even if the outgoing electron is not tagged, i.e.
 after integration over the Weizs\"acker--Williams spectrum, at this angle
 res contributions are at least two times bigger than direct ones.

 The proximity of the spectator jet from the proton  should not compromise the
 observability of this signal, since this jet is not expected to contain
 photons of
 this very high energy. Finally, due to the softness of \gga, almost the
 whole energy of the incoming photon will go into the spectator jet. If the
 energy of the photon is known (by measuring the energy of the outgoing
 electron), this information can be used to study a sample of photonic
 spectator jets with known energy, which might provide valuable information
 for the study of other resolved photon processes.

 \subsection*{4d) $J/\psi$ Production}
 Just like the production of heavy quark pairs, the process $ep \rightarrow
 e J/\psi X$ has originally been proposed \cite{oldjpsi} as a way to determine
 the gluon density of the proton;  results from such analyses
  have been reported from fixed--target photoproduction experiments
 \cite{nmcpsi}. It was realized later \cite{kust1}, however, that at the much
 higher energies which can be achieved at HERA this final state also receives
 sizable contributions from resolved photon processes. $J/\psi$ mesons can
 be produced from $\gamma g \rightarrow J/\psi\ g$ (direct process) as well
 as $gg \rightarrow J/\psi\ g$ (res process); in addition, they can be
 produced in the decay of $\chi$ mesons or $b$ quarks. Indeed, this latter
 process dominates \cite{kupsi} for $p_T > 5$ GeV; moreover, the
 cross--section becomes quite small in this region. Most studies of
 $J/\psi$ production at HERA therefore focus on the region 1 GeV $\leq p_T
 \leq$ 5 GeV.

 Although the total resolved photon contribution estimated from the DG
 parametrization amounts to about 0.5 nb, extraction of the signal may
 not be trivial. First of all, only the leptonic decays will be detectable
 at HERA, which reduces the signal by a factor of 7. Since most $J/\psi$'s
 produced via res processes emerge at large rapidity, i.e. small angle to
 the proton beam, requiring both leptons to be detected further degrades
 the signal; in this case there is no rapidity region left where the
 res contribution clearly dominates \cite{flet1}. Of course, requiring the
 spectator jet from the photon to be detected will suppress the direct
 contribution, while $J/\psi$'s produced with negative rapidity will
 overwhelmingly come from direct processes, as in case of open heavy flavour
 production.

 Another possibility is to tag the outgoing electron \cite{flet2}. This selects
 events with incident photon energy $E_{\gamma}>$ 7 GeV, since otherwise the
 outgoing electron is too energetic to be bent out of the beam, which
 is necessary for its detection in the forward spectrometer.
 \begin{figure}
 \vspace{8 cm}
 \caption{Rapidity distribution \protect\cite{flet2}
 of $J/\psi$ mesons produced at HERA in events
 where the outgoing electron is tagged, as estimated using the DO2 and DG
 parametrizations for the proton and photon, respectivley.}
 \label{psirap}
 \end{figure}
 As shown in fig.\ref{psirap},
 this is sufficient to suppress the direct contribution at positive rapidities
to an insignificant level. The price is the reduction of the overall signal
by a factor of 10 or so; on the other hand, the analysis is no longer sensitive
to the details of the spectator jet formation.

 Finally, it has been pointed out in ref.\cite{flet1} that there is no direct
 contribution if the $J/\psi$ is produced in association with a hard photon;
 this process has subsequently been studied in ref.\cite{dk2}. The main
 problem is again the small event rate; after mild acceptance cuts, the DG
 parametrization predicts an observable cross--section of only 0.08 pb.
 However, as already mentioned in connection with $J/\psi$ production at
 \eplem\ colliders, the ``colour singlet'' model \cite{berger}, which has
 been used in all cross--section calculations, might underestimate the rate
 by as much as a factor of 5 \cite{nmcpsi}. While this would make the
 detection of the signal easier, the presence of a k--factor of this
 magnitude casts doubt on the leading order analyses presented here.
 Nevertheless, $J/\psi$ production at HERA has the potential to probe
 gluon densities down to very small values of $x$, of order $10^{-3}$ or
 less.

 \subsection*{4e) Other Processes}
 We close this section with a brief survey of other processes that receive
 contributions from resolved photon reactions, although limitations of space
 do not allow to discuss them in detail.

 Of great theoretical interest is the production of $W$ and $Z$ bosons at HERA.
 In leading order only the resolved photon (Drell--Yan) process
 $q \bar{q}' \rightarrow W$ contributes \cite{dw}; note that the corresponding
 cross--section is ${\cal O}(\alpha^3/\alpas)$, since $\qvph \propto 1/\alpas$.
 On the other hand, the direct process $\gamma q \rightarrow W q'$, while
 formally of higher order in \alpas\ ($\sim \alpha^3$), is sensitive to the
 $\gamma W W$ coupling \cite{bauze}. In order to study this dependence the
 $p_T \rightarrow 0$ divergent pieces of the direct contribution, which are
 already included in the res part, have to be subtracted to avoid
 double--counting. Several subtraction procedures have been suggested recently
 \cite{wsub}. The resulting cross--section in the Standard Model
 is about 0.5 pb for $W$ bosons,
but probably only the leptonic decay mode can be identified; the cross--section
 for $Z$ bosons is even smaller.

 In subsection 4c we have discussed the production of real (on--shell) photons
 at HERA. The same processes can also give rise to off--shell photons, and
 hence to lepton pairs; this has been studied in refs.\cite{kust1,donew}.
 As expected from our previous discussions, resolved photon contributions are
 quite important at small transverse momentum and/or small invariant mass
 of the dilepton system. However, compared to direct photon production the
 cross--section is down by another factor of $\alpha$. This process can
 therefore only yield useful information about parton densities after a
 large amount of data has been accumulated.

 It has also been suggested \cite{dk1} to study the production of a photon
 whose transverse momentum is balanced by a charm quark as a means to
 constrain the heavy flavour content of the proton as well as the photon.
 The charm quark is detected via its decay into a muon. The total
cross--section
 after acceptance cuts is expected to be a few pb; the exact number
 depends on the way mass effects are included in heavy flavour density
 distributions.

 As a last process we mention the production of two hard photons at HERA
 \cite{bawa3}. The cross--section is rather small, being of order
 $\alpha^4$, so that only a limited range of transverse momenta can be
 studied experimentally. On the other hand, this process receives important
 contributions from $gg$ fusion, via a box diagram which (up to trivial
 coupling and colour factors) is equivalent to the famous light--by--light
 scattering diagram; although first studied more than 50 years ago \cite{ll},
 the effect of this diagram has still not been detected experimentally.

 \renewcommand{\theequation}{5.\arabic{equation}}
 \section*{5) Minijets and Total Cross--Sections}
 \setcounter{equation}{0}
 \setcounter{footnote}{0}
 So far we have only discussed ``hard'' processes, where the applicability of
 perturbative QCD is not in doubt. However, we have already seen in sec.3 that
 one has to introduce at least one parameter that cannot be predicted from
 perturbative QCD if one wants to describe existing \gamgam\ data in the
 intermediate region where both soft and hard processes contribute; this
 parameter is the cut--off \ptmin, which parametrizes the applicability of
 perturbative QCD. TRISTAN data indicate that this parameter has to be
 chosen around 1.5 (2.0) GeV if data are to be described by the DG (LAC1)
 parametrization. However, with such values of \ptmin, the total jet pair
 cross section grows very rapidly with energy, and eventually even exceeds the
 value of the total cross--section measured at lower energies.

 The rapid growth of the inclusive jet cross--section due to the copious
 production of ``minijets'' with $p_T \simeq \ptmin$ via resolved photon
 processes has first been pointed out in ref.\cite{dh1} for the case
 of $\gamma p$ scattering; an example is shown in fig. \ref{incjet},
 for \ptmin\ = 2 GeV and various parametrizations of \qvph.
 \begin{figure}[hbt]
 \vspace{8cm}
 \caption{Predictions \protect\cite{dh1} of the increase of the
 inclusive (mini)jet cross--section in $\gamma p$ collisions with
 \protect\rts, for \protect\ptmin\ = 2 GeV and
 various parametrizations for \protect\qvph.}
 \label{incjet}
 \end{figure}
It was conjectured in that paper that this increase of the cross--section might
 help to explain the mysteriously large number of muons observed \cite{cosmic}
 in photon--induced cosmic air showers. Later detailed Monte Carlo calculations
 \cite{cosmc} showed that, while resolved photon processes might boost the
 muon yield by a factor of 2--3, they are not sufficient to explain the data by
 themselves.

 Of course, the total cross--section cannot grow indefinitely at the rate
 shown in fig. \ref{incjet}; some mechanism will have to unitarize it.
 This problem is well known for hadronic ($pp$ or $p \bar{p}$) collisions;
 indeed, it was suggested almost 20 years ago \cite{luthe} that minijet
 production might contribute to the growth of total hadronic cross--sections.
 In this case unitarization is usually achieved by eikonalization. The
 crucial observation here is that LO QCD predictions for cross--sections,
 like those shown in fig. \ref{incjet}, refer to {\em inclusive} jet
 cross--sections; in other words, they differ from the jet production
 contribution to the total cross--section by a factor of the average jet
 pair multiplicity \nav. Formally one writes \cite{nuceik}
 \be \label{pp}
 \sigma_{pp}^{\rm inel} = \int d^2 b \left\{ 1 - exp \left[ - \left(
 \sigma_{pp}^{\rm hard}(s) + \chi_{pp}^{\rm soft} \right) A(b) \right]
 \right\},
 \ee
 Here $\vec{b}$ is the two--component impact parameter, $A(b)$ describes the
 transverse distribution of partons in nucleons, $\sigma_{pp}^{\rm hard}$
 is the perturbative QCD prediction for the minijet cross--section (obtained
 by integrating $d \sigma / d p_T$ in the region $p_T \geq \ptmin$), and
 $\chi_{pp}^{\rm soft}$ is the non--perturbative (soft) contribution to the
 eikonal, which is fitted from low--energy data. In eq.(\ref{pp}) it has
 been assumed that the transverse distribution is independent of $x$ and
 $Q^2$, and that different partonic scatterings are uncorrelated, i.e.
 obey Poisson statistics. Eikonalized minijet models with \ptmin\
 around 1.5 to 2 GeV and standard parametrizations for $\vec{q}^p$ not
 only reproduce the rise of the total and inelastic $pp$ and $p \bar{p}$
 cross--sections \cite{nuceik}, but also correctly describe many details
 of ``minimum--bias'' events as well as events containing hard jets
 \cite{nucmc}.

 However, as pointed out by Collins and Ladinsky \cite{collad},
 eq.(\ref{pp}) will have to be modified before it can be applied to
 photonic cross--sections. This can easily be seen \cite{flet3} by
 expanding the exponential; one finds that the cross--section for the
 production of 2 jet pairs is proportional to the square of the hard
 QCD cross--section. In case of $\gamma p$ scattering this hard cross--section
 is of ${\cal O}(\alpha \alpas)$, so that eq.(\ref{pp}) would predict
$\sigma$(2 pairs) $\sim {\cal O}(\alpha^2 \alpha_s^2)$. On the other hand, once
 the photon has undergone its transition into a (virtual) hadronic state,
 no additional factor of $\alpha$ is necessary to produce additional jet pairs;
 rather, one would expect $\sigma$(2 pairs) $\sim {\cal O}(\alpha \alpha_s^3)$.
 Similar arguments hold for even larger number of jet pairs. This can be
 achieved by introducting a parameter \phad\ describing the probability that
 the photon goes into a hadronic state; clearly $\phad \sim {\cal O}(\alpha)$.
 Eq.(\ref{pp}) then becomes \cite{collad}
 \be \label{gp}
 \sigma_{\gamma p}^{\rm inel} = \int d^2 b\ \phad \left\{ 1 - exp \left[ -
 \left( \sigma_{\gamma p}^{\rm hard}(s) + \chi_{\gamma p}^{\rm soft} \right)
 A(b)/\phad \right] \right\},
 \ee
 A similar expression can be derived for \gamgam\ collisions, but here
 \phad\ has to be replaced by $P^2_{\rm had}$ \cite{fs1}.

 Unfortunately there are many unknown quantities in eq.(\ref{gp}). First of
 all, we cannot predict the hard scattering cross--section, since we do not
 (yet) know the parton densities in the photon at sufficiently small values
 of $x$. TRISTAN data give some indication what \ptmin\ should be, but it
 is not clear that the same value should be used in $\gamma p$ scattering, or
 that it should be independent of energy (although first HERA data do seem to
 point in that direction). Finally, it is not clear how \phad\ and $A(b)$ are
 to be determined. In most papers \cite{collad,fs1,flet4} VMD ideas are used
 to estimate these quantities. In particular, \phad\ is taken to be the
 $\gamma \rightarrow \rho$ transition probability $\simeq 1/300$, and $A(b)$
 is computed from the Fourier transform of some pionic form factor. However,
 it should be stressed that these are {\em assumptions} which are {\em not}
 inherent to perturbative QCD or even to the idea that minijets drive the
 increase of hadronic cross--sections. Recall, for example, that in the GRV
 parametrization the ``naive'' VMD estimate of $\qvph(x,Q_0^2)$ had to be
 doubled \cite{grv} in order to describe data at higher $Q^2$. Finally,
 if one estimates \cite{flet3} \phad\ as $\int_0^1 d\!x \ x \ \qvph(x,
 p^2_{T,{\rm min}})$, one finds a value around 1/150 even for the DG
 parametrization. We therefore have to conclude that theoretical considerations
 at present only allow to estimate \phad\ up to a factor of 2 or so. The
 uncertainty in $A(b)$ has so far not been discussed in the literature, but
 might be of similar magnitude.

 In view of these ambiguities it is not surprising that predictions for
 the total $\gamma p$ cross--section at HERA energies differed quite
 widely prior to its measurement. Some examples are shown in fig. \ref{gptot},
 together with low--energy data and the recent ZEUS measurement \cite{zeustot};
 a very similar value has been reported by the H1 collaboration \cite{h1tot}.
 \begin{figure}[hbt]
 \vspace{8 cm}
 \caption{Comparison of various predictions of total $\gamma p$ cross-sections
 with low--energy data and the recent ZEUS measurement \protect\cite{zeustot}.}
 \label{gptot}
 \end{figure}
 The two solid curves show fits to low--energy data based on Pomeron
 phenomenology. The two dot--dashed curves show minijet predictions
 \cite{schuler} using the
 DG parametrization with \ptmin\ = 1.4 (upper) and 2.0 (lower curve) GeV,
 while the dotted and dashed curves have been obtained from the LAC1
 parametrization using the same values of \ptmin. The LAC parametrization
 seems disfavoured, but in view of the theoretical and experimental
 uncertainties it might be premature to exclude it altogether. The DG minijet
 prediction with \ptmin\ = 2 GeV is certainly in agreement with the data.
 Notice, however, that all minijet calculations predict a substantially
 larger slope of the cross--section than the Pomeron--based fits do; a
 measurement of the energy--dependence of the total $\gamma p$
 cross--section at HERA might therefore help to distinguish these
 models\footnote{The total cross--section can only be determined from events
 where the outgoing electron is tagged. That means that only the region
 180 GeV $\leq \rts(\gamma p) \leq 240$ GeV is available for such a
 measurement. At least in the minijet calculations the cross--section should
 vary substantially even in this limited region.}. Finally, we have already
 seen that, in the case of hadronic collisions, minijet models also reproduce
 details of event shapes, e.g. multiplicity fluctuations and various
 correlations \cite{nucmc}. The measurement of similar quantities at HERA
 should help to distinguish between models.

 Minijets are also expected to play an important role in \gamgam\ collisions
 at \eplem\ colliders \cite{dgo6,fatdg}. Indeed, the minijet cross--section
 at \eplem\ colliders rises even faster than at hadron colliders, since not
 only the \gamgam\ cross--section but also the $\gamma$ flux increases with
 energy, especially once beamstrahlung becomes important. Some examples of
 the resulting minijet cross--sections
 are shown in fig. \ref{ggtot} \cite{fatdg}, for the same photon spectra
 introduced in fig. \ref{beam}.
 \begin{figure}[hbt]
 \vspace{8 cm}
 \caption{Integrated two jet cross-section for $ p_T \geq \ptmin $
 as a function of \ptmin\ for the photon spectra of fig. \protect\ref{beam},
 as predicted \protect\cite{fatdg} from the DG parametrization.}
 \label{ggtot}
 \end{figure}
 We see that the DG parametrization with \ptmin $\simeq 1.6$ GeV predicts a
 cross--section between about 20 and 500 nb, depending
 on the machine parameters; at a \gamgam\ collider this cross--section
 would be as large as 2 $\mu$b. For the Palmer--G (Palmer--F) design
 this corresponds to about 25 (0.5) minijet pairs per bunch train
 collision; for the wbb option of the D--D and TESLA designs one expects 0.02
 and 0.004 minijet pairs, respectively, in a 100 nanosecond interval. Of
 course, the minijet cross--section is sensitive to the parton content at small
 $x$ values, where so far no experimental data exist. On the other hand,
 ``shadowing'' effects, which can be important for $x < 10^{-3}$, are not
 expected to be relevant for colliders with $\rts \leq 1$ TeV.

 We have just seen that (inclusive) jet cross--sections can be larger than
 the total cross--section. However, with the possible exception of the
 Palmer--G design whose hard beamstrahlung spectrum implies that the
 average \wgg\ and hence $\sigma(\gamgam \rightarrow$ jets) averaged over
 the photon spectrum is quite large, eikonalization effects are not
 expected to change the number of hadronic (minijet) events at 500 GeV \eplem\
 colliders significantly even if a conservative, VMD--based eikonalization
 scheme is used. This can be seen from the fact \cite{fatdg} that the
 minijet cross--section (as predicted from the DG parametrization) is smaller
 than or of the same order of magnitude as the total $\eplem \rightarrow
 \eplem +$ hadrons cross--sections estimated using a constant $\gamgam
 \rightarrow$ hadrons cross--section of 250 nb for $\wgg > 5$ GeV. We thus
 have to face the unpleasant fact that some designs for \eplem\ colliders
 predict several hadronic events to occur at each bunch train collision already
 at \rts\ = 500 GeV. It is usually accepted that a 500 GeV collider should be
 designed such that it can be upgraded to $\rts \geq 1$ TeV; of course,
 beamstrahlung and hadronic 2--photon backgrounds become worse at higher
 energies.

 This problem might be alleviated somewhat if detectors achieve a very good
 time resolution. E.g., at Palmer--F or --G, a bunch train consists of 10
 bunches in time intervals of 1.4 nsec. A time resolution of about 2 nsec
 seems achievable at least for the tracking system, so that this part of the
 detector would ``see'' at most two superimposed bunch crossings; this would
 obviously reduce the number of minijets in the smallest time unit measureable
 by the detector by a factor of 5. On the other hand, it seems unlikely that
 similarly fast calorimeters can be built. Notice that about 35 to 40\% of the
 energy of a hadronic jet is carried by neutral particles, which are only
 detectable in calorimeters.

 What are the consequeneces of ``always'' having $> 1$ minijet event present
 in the detector? Basically it means that one now has an ``underlying event'';
 i.e. every annihilation event (and every hard \gamgam\ event) will be
 accompanied by several minijet events. {\em Every} event will thus have some
 hadronic activity. This situation is of course well known from hadron
 colliders, but the absence of an underlying event, i.e. the ``cleanliness'' of
 the experimental environment, is usually considered to be one of the main
 advantages of \eplem\ colliders. The presence of a few (or even a few dozen)
 soft hadrons does usually not affect the possibility to detect ``new physics''
 signals very much, although some care has to be taken when defining what is
 meant by an isolated lepton or photon, or by a hadronically quiet event; and
 it has to be kept in mind that fluctuations in the underlying event might
 fake elements of a signal, e.g. missing $p_T$. However, the ability of
 future linacs to study new particles in detail might be compromised
 severely by the presence of a large underlying event. First of all, the beam
 energy constraint would no longer be applicable, since the visible energy can
 be larger than \rts. This already excludes the possibility of precision
 measurements of the mass of a hadronically decaying particle at energies
 far above threshold. An underlying event would also make it more difficult
 to discriminate between hadronically decaying $W$ and $Z$ bosons. Moreover, a
 large multiplicity of soft particles might make it impossible to operate a
 microvertex detector, which is deemed necessary for efficient $b$ and $c$
 quark tagging. We estimate that one minijet event will deposit between 6 and
 10 GeV of transverse energy in the detector (from both the minijets itself
 and the outer fringes of the spectator jets), corresponding to a charged
multiplicity of about 8. Finally, an underlying event would also complicate the
 study of hard \gamgam\ events, since there would always be some spectator jet
 activity in the forward and backward directions, making it much more difficult
 to distinguish between hard, direct and resolved photon events.

 It therefore seems much preferable to us to construct future \eplem\ linacs
 and their detectors such that an underlying event can be avoided. This ought
 to be relatively easy at \rts\ = 500 GeV, but might prove challenging
 \cite{fatdg} for colliders operating at $\rts \geq 1$ TeV.

 \section*{6) Summary and Conclusions}
 \begin{itemize}
 \item The measurement of \f2gam\ in deep--inelastic $e \gamma$ scattering at
 present \eplem\ colliders does not yield sufficient information for decisive
 tests of QCD, nor for a discrimination of different ans\"atze for the parton
 content of real photons (sec. 2). This is partly due to rather poor statistics
 (which is 3 or 4 orders of magnitude worse than for typical fixed--target
 deep--inelastic lepton--nucleon scattering experiments), partly due to
kinematical constraints (which do not allow measurements at small Bjorken$-x$),
 and partly because \f2gam\ is not very sensitive to \gga. The situation might
 improve at future colliders, where smaller values of $x$ become accessible
 in DIS; in this ``sea'' region, gluons do contribute to \f2gam. The ideal
 experiment of this type could be performed \cite{bawa4} if an \eplem\ linac
 can be converted into an $e \gamma$ collider by backscattering laser photons.

 \item In the last year the existence of resolved photon contributions has
 evolved from a theoretical prediction into an experimental fact. Their
 presence has first been demonstrated by the AMY group at TRISTAN, and has
 been confirmed by TOPAZ at TRISTAN and by the LEP experiments ALEPH and DELPHI
(see sec. 3). Very recently the HERA experiments H1 and ZEUS have also reported
 that their data from the first (pilot) run show clear evidence of resolved
 photon events. The three main theoretical predictions \cite{dg3} -- large
 jet cross--sections at small and moderate transverse momentum; jet rapidity
 distribution peaked at large values; and the presence of a photonic spectator
 jet -- have already been confirmed experimentally.

 \item The first analyses of resolved photon events have already contributed to
 our understanding of the hadronic structure of the photon. TRISTAN data
 clearly exclude one parametrization of photonic parton densities (LAC3); the
 measurements of the total $\gamma p$ cross--section at HERA are in conflict
 with predictions from the more extreme variety of minijet models. TRISTAN and
 LEP data will improve due to increased statistics, improved angular coverage
 of the detectors (at TRISTAN) and increased beam energy with less annihilation
 backgrounds (at LEP). The next year should see the HERA data sample grow by
at least 3 orders of magnitude. The number of resolved photon events detected
at
 HERA will then greatly exceed that of all \eplem\ colliders combined, allowing
 for detailed studies of jet production as well as searches for many other
 final states (sec.4). Nevertheless, \eplem\ data will continue to play an
 important role. On the one hand, these lower energy (in the \gamgam\ or
 $\gamma p$ centre--of--mass system) machines can probe the parton densities in
 the photon at large $x$ but moderate $Q^2$, while at higher energies large
 $x$ usually imply large $Q^2$. Recall that all models converge towards the
 asymptotic prediction if both $x$ and $Q^2$ are large, while there are
 sizable differences at large $x$ and moderate $Q^2$. Moreover, \eplem\
 colliders also allow to study events with rather small invariant mass, which
 are usually boosted out of the detector at HERA; this should help us in
 understanding the transition between soft and hard interactions.

 \item Soft and semihard (minijet) \gamgam\ events can lead to an
 ``underlying event'' at future \eplem\ supercolliders, spoiling the
 traditional cleanliness of \eplem\ colliders (sec. 5). The main question
 here is whether beamstrahlung can be kept under control. Existing designs
 indicate that this should be fairly easy at centre--of--mass energies
 up to 500 GeV, but can become increasingly difficult at higher energies.
 \end{itemize}

 Our general conclusion is that the importance of resolved photon contributions
 increases with beam energy, and thus with time. We therefore expect great
 progress to be made in this field over the next few years. This is the
 heroic age of resolved photons!

 \subsection*{Acknowledgements}
 We thank our experimental colleagues, G. d'Agostini from ZEUS, A. Finch from
 ALEPH, H. Hayashii from TOPAZ and T. Nozaki from AMY, for keeping us informed
 about the latest stage of the analyses of their data. M.D. thanks the
 theory group of the TIFR for their kind hospitality during his stay.

 \end{document}